\documentclass[fleqn,usenatbib]{mnras}
\usepackage{newtxtext,newtxmath}
\usepackage[T1]{fontenc}
\DeclareRobustCommand{\VAN}[3]{#2}
\let\VANthebibliography\thebibliography
\def\thebibliography{\DeclareRobustCommand{\VAN}[3]{##3}\VANthebibliography}
\usepackage{graphicx}
\usepackage{amsmath,subfig}
\usepackage{tcolorbox}
\usepackage{hyperref}

\def\Fig{\mbox{Figure~}}
\def\Figs{\mbox{Figures~}}
\def\Tab{\mbox{Table~}}

\def\Sec{\mbox{Section~}}
\def\Secs{\mbox{Sections~}}
\def\Eq{\mbox{Equation~}}

\def\Re{\mbox{$R_{\rm e}$}}

\title[LeMoN]{\textsc{LeMoN}: Lens Modelling with Neural networks -- I. Automated modelling of strong gravitational lenses with Bayesian Neural Networks}

\author[F. Gentile et al.]
{Fabrizio Gentile,$^{1,2}$\thanks{E-mail: fabrizio.gentile3@unibo.it}
Crescenzo Tortora,$^{3}$
Giovanni Covone,$^{3,4,5}$
Léon V.E. Koopmans,$^{6}$
Rui Li,$^{7}$
\newauthor
Laura Leuzzi$^{1,2}$
and Nicola R. Napolitano$^{3,7}$
\\
$^{1}$University of Bologna, Department of Physics and Astronomy (DIFA), Via Gobetti 93/2, I-40129, Bologna, Italy\\
$^{2}$INAF -- Osservatorio di Astrofisica e Scienza dello Spazio, via Gobetti 93/3 - 40129, Bologna - Italy\\
$^{3}$INAF -- Osservatorio Astronomico di
Capodimonte, Salita Moiariello, 16, 80131 - Napoli, Italy\\
$^{4}$Dipartimento di Fisica "Ettore Pancini",
Universit\`{a} di Napoli Federico II, Compl. Univ. Monte S.
Angelo, 80126 - Napoli, Italy\\
$^{5}$INFN, Sezione di Napoli, C.U. Monte S. Angelo, Via Cinthia, I-80126 Napoli, Italy\\
$^{6}$Kapteyn Astronomical Institute, University of Groningen, P.O Box 800, 9700 AV Groningen, The Netherlands \\
$^{7}$School of Physics and Astronomy, Sun Yat-sen University Zhuhai Campus, Daxue Road 2, 519082—Tangjia, Zhuhai, Guangdong, China
}

\date{Accepted XXX. Received YYY; in original form ZZZ}

\pubyear{2022}

\begin{document}
\label{firstpage}
\pagerange{\pageref{firstpage}--\pageref{lastpage}}
\maketitle

\begin{abstract}
The unprecedented number of gravitational lenses expected from new-generation facilities such as the ESA \textit{Euclid} telescope and the \textit{Vera Rubin Observatory} makes it crucial to rethink our classical approach to lens-modelling. In this paper, we present \textsc{LeMoN} (Lens Modelling with Neural networks): a new machine-learning algorithm able to analyse hundreds of thousands of gravitational lenses in a reasonable amount of time. The algorithm is based on a \textit{Bayesian Neural Network}: a new generation of neural networks able to associate a reliable confidence interval to each predicted parameter. We train the algorithm to predict the three main parameters of the Singular Isothermal Ellipsoid model (the Einstein radius and the two components of the ellipticity) by employing two simulated datasets built to resemble the imaging capabilities of the \textit{Hubble Space Telescope} and the forthcoming \textit{Euclid} satellite. In this work, we assess the accuracy of the algorithm and the reliability of the estimated uncertainties by applying the network to several simulated datasets of $10^4$ images each. We obtain accuracies comparable to previous studies present in the current literature and an average modelling time of just $\sim$0.5s per lens. Finally, we apply the \textsc{LeMoN} algorithm to a pilot dataset of real lenses observed with HST during the SLACS program, obtaining unbiased estimates of their SIE parameters. The code is publicly available on GitHub (\url{https://github.com/fab-gentile/LeMoN}).
\end{abstract}

\begin{keywords}
gravitational lensing: strong -- galaxies: elliptical and lenticular, cD -- methods: data analysis -- software: data analysis 
\end{keywords}



\section{Introduction}
Initially predicted by Albert Einstein as a consequence of his General Relativity \citep{einstein_1915}, gravitational lensing consists in the deflection of light caused by a gravitational field. This phenomenon can be very intense when a massive object (e.g., a galaxy or a cluster) is involved. In this case, lensing can produce multiple images of a distant compact source \citep[see e.g.][]{Huchra_1985,Napolitano_2020} or gravitational arcs can be formed in case of extended sources \citep{Zwicky_Lensing}. In both these cases, the phenomenon is called “strong gravitational lensing”, while the composite systems created are generally known as “gravitational lenses”. 
The main observables related to this phenomenon (i.e., the position and shape of the lensed images) mainly rely on the matter distribution inside the lensing galaxy and on the relative \textit{angular-diameter distances} involving the observer, the deflector, and the background source \citep[e.g.,][]{schneider_1992,bartelmann_theory}. An additional contribution (up to 15\%; see e.g. \citealt{schneider_1992}) can also derive from the line-of-sight matter distribution. Strong lensing can be successfully employed to study the dark matter distribution in galaxies \citep[e.g.][]{treu_DM,Covone_2009,Tortora_DM,Auger_DM,spiniello_DM,Sonnenfeld_15,Shajib_21} and - since the distances involved in lensing are sensitive to the cosmological parameters - to measure the Hubble constant with the so-called “time-delay technique” \citep[e.g.,][]{refsdal_hubble,Wong_Holicow2}. Further interesting applications of lensing span from identifying dark matter substructures in galaxies \citep[e.g.,][]{mao_substructure,dalal_substructure,Koopmans_DM,Vegetti_DM} to studying the universality of the Initial Mass Function \citep[e.g.,][]{Treu_IMF,Auger_IMF,barnabe_IMF,Sonnenfeld_IMF}. A more comprehensive review of the possible scientific applications of strong lensing in modern astrophysics and cosmology can be found in \citet{Treu_Lensing} and \citet{Congdon_Lensing}. 

For decades, however, the main limitation to the scientific potentialities of strong lensing has been represented by the small number of known and analysed gravitational lenses. The need for massive objects acting as lenses and an almost perfect alignment between observer, foreground lens, and background source make gravitational lenses rare \citep[e.g.][]{schneider_1992}. Several studies estimated that less than one galaxy out of $10^{3-4}$ shows detectable lensing features \citep[see e.g.][]{Collett_LensPop}. Therefore, the inspection of large samples of astronomical sources is needed to identify a statistically significant sample of lenses.

This scenario, however, is about to change dramatically in the near future. A new generation of telescopes such as the ESA \textit{Euclid} satellite \citep{Laureijs_Euclid} and the \textit{Vera Rubin Observatory} \citep{LSST} will be soon on the scene. Several forecasts claimed how these facilities will provide data on more than one billion galaxies, allowing the identification of $\sim10^5$ new homogeneously selected gravitational lenses \citep[see e.g.][]{Collett_LensPop,Laureijs_Euclid,LSST,Serjeant_Lenses}. These newly-discovered systems will increase by several orders of magnitude the number of known gravitational lenses, allowing many astrophysical studies to rely on a stronger statistics, and to explore a wider range of redshifts and galaxy properties. Moreover, a whole new approach to lensing-based science - the so-called "statistical lensing" - will be possible \citep[e.g.][]{sonnenfeld_2021a,sonnenfeld_2021b,sonnenfeld_2021c,sonnenfeld_2021d}.

Although these are important advances, the massive amount of data produced by these instruments would to overwhelm our ability to analyse them with classical techniques. For this reason, significant effort has been afforded in the last years to rapidly search for gravitational lenses in vast datasets, mainly thanks to the employment of machine learning algorithms (see, e.g., the noteworthy results of the first \textit{strong gravitational lens finding challenge} in \citealt{Metcalf_Challenge}). Several sky surveys have been systematically analysed with these techniques, allowing the identification of an unprecedented number of likely lensed objects \citep[see some examples in][]{Petrillo_1,Petrillo_2,Petrillo_3,Li_KiDS,He_KiDS,Jacobs_DES,Jacobs_DES2,Canameras_HoliSmokes,Gentile_2022}. Nevertheless, our ability to model lenses – necessary to allow the scientific exploitation of the retrieved systems - remained almost unchanged in the required compute time. Currently, most lens-modelling algorithms rely on Bayesian analysis, such as \textit{Monte Carlo Markov Chains} or \textit{nested sampling} (see some noteworthy examples in \citealt{Jullo_2007,Vegetti_2008,Birrer_2018,Nightingale_2018} and \citealt{Lefor_2013} for a comparison between the different methods). These techniques are computationally expensive (the analysis of a single lens can require up to several hours) and often require a non-trivial human intervention to select the lensing feature in the image. These properties make them less suitable to efficiently model large samples of lenses. As it happened for the search for strong lenses, an efficient approach to lens-modelling can come from machine-learning. Several algorithms have been proposed. These are mainly based on \textit{Convolutional Neural Networks} (CNNs; see e.g. \citealt{Hezaveh_CNN,pearson_ModellingCNN,Schuldt_ModellingCNN}) or \textit{Bayesian Neural Networks} (BNNs; e.g. \citealt{Levasseur_modellingBNN,bom_2019,Schuldt_22}): supervised-learning algorithms able to extract meaningful features from images and to convert them into useful parameters \citep{LeCun_CNNReview,Charnock_BNN}. In this work, we focus on BNNs. This new generation of machine learning algorithms is able to account, in its predictions, for the different sources of uncertainty that can affect the analysed data and the algorithm itself, providing a full Bayesian treatment of them. Thanks to this property, BNNs can associate reliable confidence intervals to the estimated parameters, allowing a wide range of scientific applications (see some astrophysical examples in \citealt{Cobb_2019,Hortua_CMB,Rivera_2020}).

In this paper, we propose a new lens-modelling algorithm called “\textsc{LeMoN}”\footnote{\url{https://github.com/fab-gentile/LeMoN}} (LEns MOdelling with Neural Networks). Based on a BNN, it can efficiently model large samples of lenses in a reasonable amount of time. This work - the first of the \textsc{LeMoN} series - presents the algorithm and its training, assessing its performances in modelling both simulations and a pilot sample of real lenses. The paper is structured as follows: \Sec\ref{sec:data} presents the datasets employed to train and test the algorithm. \Sec\ref{sec:methods} introduces BNNs, and the architecture implemented in the \textsc{LeMoN} algorithm. In \Sec\ref{sec:results}, we assess the algorithm’s performance on a simulated dataset and on a small set of real lenses from the \textit{Sloan Lens ACS Survey} (SLACS; \citealt{Slacs_I}). \Sec\ref{sec:discussion} discusses the results obtained on all the datasets and measures the performance on the estimation of different lens parameters. Finally, in \Sec\ref{sec:conclusions}, we draw our conclusions and list the future perspective of this work.

\begin{figure}
	\includegraphics[width=\columnwidth]{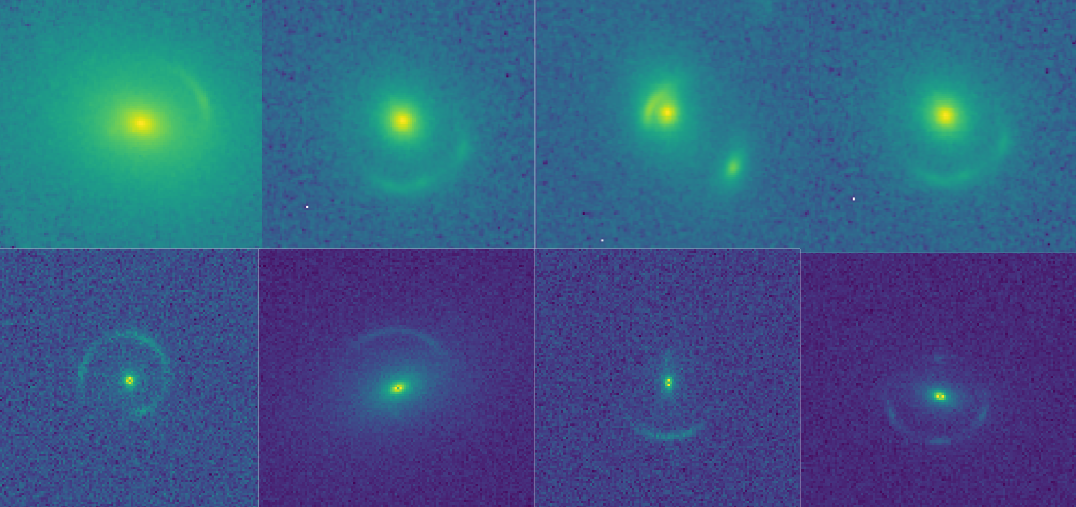}
    \caption{Some examples of the mock lenses employed in this work. The images in the top row are \textit{HST mocks} and are simulated to recall the imaging capabilities of the \textit{Advanced Camera for Surveys} (ACS) mounted on the Hubble Space Telescope. The images in the bottom row are \textit{Euclid mocks} and mimic the observational characteristics of the systems observed with the forthcoming VIS imager of the \textit{Euclid} telescope. Further details in \Sec\ref{sec:data}.}
    \label{fig:Mocks}
\end{figure}

\section{Building the training set}
\label{sec:data}

Training a supervised-learning algorithm, such as a BNN, consists in passing large numbers of “labelled” examples (i.e., images for which the ground truth is known) to the network. When dealing with strong gravitational lensing, this task is particularly challenging because of the small number of known and modelled gravitational lenses. The largest homogeneously selected collections of confirmed lenses \citep[e.g.][]{Slacs_I,SL2S_first,SL2S_last,Slacs_XIII} contain only a few hundreds objects: at least two orders of magnitude less than the number of examples required to find the optimal value for the millions of free parameters inside a BNN (see e.g. the sizes of the datasets employed by \citealt{Levasseur_modellingBNN} or \citealt{bom_2019}). In addition, these samples do not homogeneously cover all possible lensing configurations, being generally biased towards systems with more prominent lensing features. Since supervised-learning algorithms mainly act as interpolators (i.e. they are sensitive only in the range of parameters well covered by the training set; see  e.g. \citealt{Goodfellow_ML}), often these samples are not well suited as training sets. For the above reasons, most studies focused on the search and modelling of gravitational lenses with machine-learning methods rely on simulated data. 

In this work, we simulate two different training sets. The first one is composed by \textit{Euclid mocks}: it is built to resemble the observational characteristics expected from the \textit{Euclid}'s VIS imager as forecasted by \citet{Laureijs_Euclid} and \citet{Cropper_VIS}. The second training set (composed by the \textit{HST mocks}) mimics the imaging capabilities of the \textit{Advanced Camera for Surveys} (ACS; \citealt{Ryon_ACS}) mounted on the \textit{Hubble Space Telescope} (HST). Both the datasets are composed by 40.000 simulated images each. Some examples of these images are reported in \Fig\ref{fig:Mocks}. We underline that in this work we assume that is possible to retrieve a completely pure sample of strong lenses (without any contaminant, quite common in samples selected through visual inspection or machine-learning techniques; see, e.g., \citealt{Metcalf_Challenge,Petrillo_1,Petrillo_2,Petrillo_3} and \citealt{Gentile_2022}). The impact of contaminants on the algorithm's accuracy will be investigated in the forthcoming papers of the \textsc{LeMoN} series.

\subsection{Simulating gravitational arcs} 
\label{sec:data_Arcs}

Both simulation procedures start with a set of mock gravitational arcs. These are generated from scratch through the simulation code presented in \citet{Chatterjee_PhD} and previously employed in several studies on strong lensing \citep{Petrillo_1,Petrillo_2,Petrillo_3,Gentile_2022}. The code firstly simulates a high-\textit{z} galaxy through a Sérsic profile \citep{Sersic_Profile}, whose parameters are randomly sampled from the distributions reported in \Tab\ref{tab:param}. Uniform sampling is adopted for three parameters (axis ratio, position angle and Sérsic index), while the effective radius is sampled from a logarithmic distribution. This choice is necessary to account for the likely smaller size of high-\textit{z} sources, as discussed in \citet{Petrillo_1}. The code also adds a random number of Sérsic components (up to five) to the main light distribution to crudely mimic star-forming regions in the lensed galaxy \citep{Petrillo_2}. The parameters of these components are sampled from the distributions reported in \Tab\ref{tab:param}.

Once the background source is generated, the code simulates the foreground deflector’s matter distribution through a Singular Isothermal Ellipsoid model (SIE; \citealt{Kormann_SIE}). All the model parameters are uniformly sampled from the distributions in \Tab\ref{tab:param}. It is important to underline that, in order to save memory during the training phase, we sample the Einstein radii for the \textit{HST mocks} from a smaller distribution than for the \textit{Euclid mocks}. This step is crucial to prevent the most asymmetrical arcs from being cut out from the stamp due to the better resolution of the HST detector.  To increase the complexity of the lensing galaxy and to account for the likely presence of a matter distribution along the line-of-sight, the code also adds a Gaussian Random Field to the gravitational field generated by the SIE model. More details on this step and the parameters employed in the simulation (i.e. the form of the power-spectrum and the variance of the field) can be found in \citet{Hezaveh_16} and \citet{Petrillo_2}

Once simulated both the background source and the foreground deflector, the code performs the ray-tracing and generates the gravitational arc. For doing so, the software randomly poses the lensed source within a distance given by the tangential caustics of the SIE plus one effective radius of the source Sérsic profile from the centre of the model. This step is needed to populate the training set with a higher percentage of highly-magnified arcs with respect to lensed systems with less-magnified two or four lensed images. Simulations are performed on a 121x121 pixel grid for the \textit{Euclid mocks} (corresponding to a 12 arcsec side at the 0.1 arcsec pixel$^{-1}$ resolution expected from the \textit{Euclid} VIS imager; \citealt{Cropper_VIS}). Similarly, the simulations of \textit{HST mocks} are performed on a 131x131 grid (corresponding to a 3.9 arcsec side at the 0.03 arcsec pixel$^{-1}$ resolution of the ACS camera; \citealt{Ryon_ACS}). 

The final step of our simulation procedure consists of a convolution with a Gaussian 2D-kernel. This step produces more realistic images by accounting for the PSF blurring. The Full Width at Half Maximum (FWHM) values of the kernels are chosen as 0.16 and 0.08 arcsec to mimic the PSF expected from the \textit{Euclid} and HST images, respectively \citep{Ryon_ACS,Cropper_VIS}. 

\begin{table}
\centering
\caption{Distribution employed in \Sec\ref{sec:data} to randomly sample the parameters used during the simulations of the different training sets. All the parameters are uniformly sampled, except where indicated. Further details in \Sec\ref{sec:data}.}
\begin{tabular}{ccc}
\hline 
Parameter & Range & Units  \\
\hline 
 \multicolumn{3}{c}{Lens (SIE)}  \\
\hline 
Einstein radius$^{(a)}$ & 0.5 - 3.0& arcsec  \\
Axis ratio & 0.3 - 1.0 & -  \\
Major-axis angle & -90 - 90 & degrees  \\
\hline
 \multicolumn{3}{c}{Source (Sérsic)} \\
\hline 
Effective Radius (\Re) & 0.2 - 0.6 (log$_{10}$) & arcsec  \\
Axis ratio & 0.3 - 1.0 & -  \\
Major-axis angle & -90 - 90 & degree  \\
Sérsic Index & 0.5 - 5.0 & -  \\
\hline 
\multicolumn{3}{c}{Sérsic Blobs (1 up to 5)}  \\
\hline 
Effective radius & $(1\% - 10\%) \Re$ & arcsec  \\
Axis ratio & 1.0 & -  \\
Major-axis angle & 0 & degrees  \\
Sérsic Index & 0.5 - 5.0 & -  \\
\hline 
\end{tabular} 
\begin{flushleft}
{\footnotesize $^{\mathrm (a)}$ The Einstein radii for the \textit{HST-like} datasets are sampled from a smaller distribution with maximum value 2 arcsec to account for the different resolution of the instrument and avoid gravitational arcs escaping from the cutouts. }
\end{flushleft}
\label{tab:param}
\end{table}

\subsection{Simulating the lens light}
\label{sec:data_lens}

A realistic image of a gravitational lens must include the light distribution of the foreground deflector. Moreover, since the possible blending between the arc and the central galaxy can make it harder to detect the lensing features, it can sensibly affect the modelling accuracy of our algorithm. In the current literature, two different strategies have been followed to simulate the deflectors. In some studies \citep[e.g.][]{Pourrahmani_LensFlow,Metcalf_Challenge}, the deflector is generated from scratch through an analytical brightness profile. In others, mock gravitational arcs are stacked on real galaxy images employed as likely deflectors \citep[e.g.][]{Petrillo_1,Gentile_2022}. In this work, we employ both strategies. 

\subsubsection{HST mocks} 

For the \textit{HST mocks}, we employ as deflectors real galaxies observed in the \textit{Cosmic Evolution Survey} (COSMOS) field by the Hubble Space Telescope \citep{Koekemoer_2007,Massey_2010}. Firstly, we select the brightest galaxies in the COSMOS2020 catalogue\footnote{\url{https://cosmos2020.calet.org/}} \citep{Weaver_2021}, collecting all the sources with a F814W magnitude brighter than 21.5 \citep{Gentile_2022}. Among these sources, we select the early-type galaxies, which are known to represent the majority of the gravitational lenses \citep[e.g.][]{Eisenstein_LRG,Oguri_LRG}. We employ the “\textsc{Farmer}” version of the COSMOS2020 catalogue to perform this second selection. Since this version contains the photometry extracted through the profile-fitting code “\textsc{The Farmer}” (Weaver et al, in prep), it also includes a value named "\textsc{Solution\_Model}" describing the best-fitting brightness profile for each source. We select all the galaxies with \textsc{Solution\_Model} equal to "\textit{DevGalaxy}" (i.e. galaxies with a de Vaucouleurs brightness profile; \citealt{deVaucouleurs_Profile}) or "\textit{SimpleGalaxy}" (partially resolved galaxies with a circular exponential profile). In doing so, we collect a sample of $\sim4.000$ likely early-type galaxies and retrieve HST imaging for these sources from the public COSMOS archive\footnote{\url{https://irsa.ipac.caltech.edu/data/COSMOS/}}. A visual inspection of a significant fraction of these sources confirmed a contamination rate (i.e. a fraction of spirals and irregular galaxies) lower than 10\%. Since this tiny percentage of contaminants is not expected to sensibly affect the training, these images are not removed from the dataset.

To finally simulate \textit{HST mocks}, we follow a slightly modified version of the strategy employed in \citet{Petrillo_1,Petrillo_2,Petrillo_3} and \citet{Gentile_2022}:
\begin{enumerate}

\item We simulate a gravitational arc, as described in \Sec\ref{sec:data_Arcs}; 

\item We randomly select a galaxy from the aforementioned sample of likely deflectors. To avoid excessive contamination from the central galaxy light, we choose only galaxies with an FWHM\footnote{The value of the FWHM is the \textsc{FWHM\_F814W} entry in the COSMOS2020 catalogue, obtained from the HST imaging.} in the range $[0.1,1.1]\theta_E$. These values are broadly consistent with the findings by \citet{SLACS_5};

\item We randomly rotate the deflector image by an angle in the range $[0,360]$ degrees and flip it on the horizontal axis with a probability of 50\%;

\item We stack the two images and rescale the brightness of the gravitational arc to $\alpha B$, where $B$ is the maximum brightness of the central galaxy and the $\alpha$-factor is uniformly sampled in the range $[0.03,0.2]$. The $\alpha$-factor accounts for the typical brightness ratio between lenses and arcs \citep[e.g][]{Petrillo_1}. 

\item We perform a square-root stretching of the co-added image to enhance the low-brightness lensing features; 

\end{enumerate}

We underline that - by employing real galaxies as deflectors - we do not need to simulate the background noise or add additional sources in the lens environment to account for the likely presence of nearby galaxies. 

\subsubsection{Euclid mocks} 

The procedure exposed in the previous paragraph cannot be employed to simulate \textit{Euclid mocks}. Since – at the moment – we do not have data on real galaxies observed by the \textit{Euclid} satellite, we simulate from scratch the light distribution of the deflector by employing a Sérsic profile \citep{Sersic_Profile}. The forthcoming papers of the \textsc{LeMoN} series will use a more accurate method by employing more realistic deflectors. The brightness profile's axis ratio and position angle values are uniformly sampled from the same distributions employed in \Sec\ref{sec:data_Arcs} and summarised in \Tab\ref{tab:param}. The Sérsic index is fixed to the value $n=4$ to account for the typical elliptical galaxies in the gravitational lenses population \citep[e.g.][]{Eisenstein_LRG,Oguri_LRG}. The effective radius is uniformly sampled from the distribution $[0.1,1.1]\theta_E$ to obtain images similar to those produced for the \textit{HST mocks}.  

Differently from the \textit{HST mocks}, since the deflectors are generated from scratch, we also need to simulate the background noise. This is done through a Gaussian random-number generator: the mean value and the standard deviation of this normal distribution are computed starting from the average sky brightness (22.2 mag) and exposure time (1.610s) expected from the \textit{Euclid wide survey }\citep{Laureijs_Euclid,Scaramella_Euclid_WS}. Once a noise map is simulated, we rescale the brightness of the arc to obtain an integrated SNR in the range $[5,20]$ and rescale the brightness of the central galaxy to get an $\alpha$-factor in the same range employed in the previous paragraph.

\section{Bayesian Neural Networks}
\label{sec:methods}

In a computational perspective, the problem of lens-modelling can be reduced to assigning a set of continuous values (i.e. the set of parameters describing the mass distribution in the lensing galaxy) to an image. From a machine-learning point of view, this task is a regression problem and - therefore - it can be successfully addressed with supervised-learning algorithms such as \textit{Convolutional Neural Networks} (CNNs; e.g. \citealt{LeCun_CNNReview}). These algorithms can extract the most relevant features from an image and - once provided a proper training set - can approximate the complex relationship between the input images and the target values. For this reason, CNNs are nowadays employed in many astrophysical studies spanning from the morphological classification of galaxies \citep[e.g.][]{Dielman_Morphology,Aniyan_Morphology,Dominguez_Morphology} to cosmological studies \citep[e.g.][]{Fluri_Cosmology,Canameras_HoliSmokes}.
Nevertheless, in the last years, some studies started to highlight several limits in the use of CNNs (see a review of the main ones in \citealt{Kendall_BNN} and \citealt{Charnock_BNN}). Among these, the difficulty to associate a confidence interval to the predicted parameters has probably the most significant impact on the scientific potentialities of these algorithms. Estimating the uncertainties affecting the scientific quantities is generally required to compare them with the theoretical predictions. Moreover, a confidence interval is generally useful in order to exclude statistical outliers (in our case, the systems for which the lens modelling is not reliable). A possible way to overcome this problem and obtain an estimation of the uncertainties is represented by \textit{Bayesian Neural Networks} (BNNs; e.g. \citealt{Charnock_BNN}). This new generation of machine-learning algorithms is built by starting from a classical neural network (or a CNN when the analysis of images is needed) but it can also provide a fully Bayesian treatment of the different sources of uncertainty affecting the input data and the algorithm itself. Even though the concept itself of uncertainty is quite debated, especially in the field of machine-learning, there is an almost unanimous consensus about what a BNN is required to account for in its predictions. It mainly consists in what the algorithm "does not know" (the so-called \textit{epistemic uncertainty}) and what it "cannot know" (i.e. the \textit{aleatoric uncertainty}).

\subsection{Epistemic Uncertainty}
\label{sec:methods_epistemic}
The first kind of uncertainty that we have to model concerns all the features that a BNN cannot recognise because of a lack of proper training. Since - as discussed in \Sec\ref{sec:data} - we do not expect our algorithm to be sensitive on ranges of parameters insufficiently well-covered by our training set, we expect the predictions on images "too different" from those on which we trained the algorithm to be "more uncertain". To translate this qualitative concept into a confidence interval, we have to pose the problem of uncertainty into a Bayesian framework.

It can be done by representing a "standard" (convolutional) neural network as a highly complex parametric function
\begin{equation}
    \mathbf{y}=f(\mathbf{x},\mathbf{\omega}),
\end{equation}
providing a set of output values ($\mathbf{y}$) for each input datum ($\mathbf{x}$), once provided the weights ($\mathbf{\omega}$). Moreover, since the weights are adjusted during the training, they can be expressed as an (unknown) function of the data in the training set ($\mathbf{X}$) and their target values ($\mathbf{Y}$):
\begin{equation}
\mathbf{\omega}=\mathbf{\omega}(\mathbf{X},\mathbf{Y})
\end{equation}
In a Bayesian probabilistic framework, the uncertainty on the output values can be evaluated through their posterior probability distribution:
\begin{equation}
\label{eq:posterior}
   p(\mathbf{y}) = p(\mathbf{y}|\mathbf{x},\mathbf{\omega})
\end{equation}
In principle, this term could be evaluated by marginalising over all the possible values of the weights.
\begin{equation}
\label{eq:integral}
    p(\mathbf{y}|\mathbf{x},\mathbf{X},\mathbf{Y})=\int_\Omega p(\mathbf{y}|\mathbf{x},\mathbf{\omega}) p(\mathbf{\omega}|\mathbf{X},\mathbf{Y}) \, d\mathbf{\omega}
\end{equation}
However, due to the high dimensionality of the weights space $\Omega$ and the lack of knowledge about the posterior probability of the weights $p(\mathbf{\omega}|\mathbf{X},\mathbf{Y})$, \Eq\ref{eq:integral} cannot be evaluated in a computationally affordable way. A possible solution to this problem can reside in variational inference \citep[e.g.][]{Jordan_Variational}. This technique consists of approximating the unknown posterior with an analytic and parametric function $q_\mathbf{\theta}(\mathbf{\omega})$ with $\mathbf{\theta}$ variational parameters by minimising the difference between these distributions through the Kullback–Leibler (KL) divergence \citep{Kullback_Divergence}. In this way, \Eq\ref{eq:integral} can be rewritten as 
\begin{equation}
\label{eq:variational}
p(\mathbf{y}|\mathbf{x})\approx\int p(\mathbf{y}|\mathbf{x},\mathbf{\omega}) q_\mathbf{\theta}(\mathbf{\omega}) \, d\mathbf{\omega}    
\end{equation}
allowing a more straightforward evaluation of the probability distribution on the output value. It can be shown \citep[see e.g.][]{Gal_2015,Gal_2016} that minimising the KL divergence is equivalent to maximising the log-evidence lower bound with respect to the variational parameters $\mathbf{\theta}$:
\begin{equation}
\label{eq:loss_ori}
    \mathcal{L}_{VI}=\int q_\mathbf{\theta}(\mathbf{\omega}) \log p(\mathbf{Y}|\mathbf{X},\mathbf{\omega}) \, d\mathbf{\omega} - KL(q_\mathbf{\theta}(\mathbf{\omega})||p(\mathbf{\omega}))
\end{equation}
A possible choice for the variational function $q_\mathbf{\theta}(\mathbf{\omega})$ is the one employed in the so-called “\textit{Monte-Carlo dropout}” technique \citep{Gal_2016}:
\begin{equation}
\label{eq:choice}
    \begin{cases}
    \mathbf{\omega}_i=\mathbf{\theta}_i \cdot \rm{diag}(\{s_{ij}\}_{j=1}^{i-1}) \\
    s_{ij}=\rm{Bernoulli}(p_i)
    \end{cases}
\end{equation}
where $s_{ij}$ are Bernoulli-distributed random variables with probability $p_i$. With this choice, the first term in \Eq\ref{eq:loss_ori} becomes the log-likelihood of the network’s predictions, while the second term can be approximated with an $L_2$ regularisation with a $\lambda$ parameter \citep{Gal_2015}:
\begin{equation}
\label{eq:loss}
    \mathcal{L}_{VI}\sim\sum_{i=1}^N\mathcal{L}(\mathbf{y}_i,\mathbf{y}_i^\star(\mathbf{x}_i,\mathbf{\omega}))-\lambda\sum_j||\mathbf{\omega}_j||^2
\end{equation}

Where $\mathcal{L}$ is the log-likelihood for the network's predictions $\mathbf{y}_i^\star$ on the input datum $\mathbf{x}_i$ with real values $\mathbf{y}_i$ and weights $\mathbf{\omega}$ randomly sampled from the variational distribution $q_\mathbf{\theta}(\mathbf{\omega})$.
Moreover, the choice in \Eq\ref{eq:choice} allows us to evaluate the integral in \Eq\ref{eq:variational} with a simple Monte Carlo integration. Specifically, it can be easily implemented with the "\textit{dropout}" technique \citep{Srivastava_Dropout} consisting of randomly switching off some connections between the neurons in the neural network with a “\textit{p}” probability ($1-p$ is the so-called \textit{keep-rate}). The dropout is implemented both at training time and at testing time. In doing so, each prediction made on the same input datum with some connections randomly dropped out is equivalent to a sampling from the posterior probability presented in \Eq\ref{eq:posterior}. With a sufficiently high number of predictions, we can reconstruct the probability distribution and, therefore, evaluate its “epistemic” uncertainty by measuring its standard deviation.

\subsection{Aleatoric Uncertainty}
\label{sec:methods_aleatoric}

A second kind of uncertainty that is crucial to evaluate properly is the so-called “aleatoric uncertainty”. This value is mainly related to the intrinsic quality of the data analysed by the algorithm. Corruptions in the images (e.g. the presence of a masked region), PSF blurring, source blending and a low value of the SNR are among the most common sources of aleatoric uncertainty. It is noteworthy to underline that - differently from the epistemic uncertainty that, in principle, could be decreased by employing a more complete training set - the aleatoric uncertainty only depends on the quality of the analysed data \citep[see e.g][]{Charnock_BNN,Kendall_BNN}. Since it is a function of the only input data, aleatoric uncertainty must be evaluated for each input image separately. This task is however challenging, since these values are not available for the images in the training set. Therefore, aleatoric uncertainties must be evaluated in an unsupervised framework. A common choice (but not the only one; see e.g. Fagin et al. in prep) to estimate this quantity consists in employing a \textit{Gaussian negative-log-likelihood} as the loss function to be minimised during the training (\Eq\ref{eq:loss}):
\begin{equation}
\label{eq:loss_ale}
    \log\mathcal{L}(\mathbf{y}_i,\mathbf{y}_i^\star(\mathbf{x}_i,\mathbf{\omega}))=\sum_{j=1}^N \left[-\frac{1}{2\sigma_{i,j}^2}||y_{i,j}-y^\star_{i,j}|| -\frac{1}{2}\log(\sigma_{i,j}^2)\right]
\end{equation}
Where $\sigma_{i,j}$ is the aleatoric uncertainty predicted for the $j$-th parameter of the $i$-th image. We underline that the second term in \Eq\ref{eq:loss_ale} prevents the algorithm to predict an infinite uncertainty for all the images, regardless of the input.

\subsection{\textsc{LeMoN} Architecture}
\label{sec:methods_lemon}

In this paper, we present \textsc{LeMoN} (LEns MOdelling with Neural networks): a lens-modelling algorithm based on the Bayesian Neural Network described in the previous subsection. Our network is able to model both the aleatoric and epistemic uncertainties affecting the data and the training process and to combine them into a single confidence interval for each predicted parameter.

We implement our BNN starting from a standard CNN with a \textit{ResNet-34} architecture \citep{He_ResNet}. The whole code is written in \textsc{Python 3.9}, employing the open-source library \textsc{Keras} \citep{Chollet_keras} with a \textsc{TensorFlow} back-end \citep{Abadi_TensorFlow}. As described in the previous sections, we modify the architecture of the CNN by adding a dropout layer after each weight layer both in the convolutional blocks and in the final \textit{fully-connected} layer. While the first layer is designed to take as input a single-band image of a gravitational lens, the last layer of the network (i.e. the one that performs the regression) is composed of six nodes.  Three of them will predict the three parameters of the SIE model (\Sec\ref{sec:data}) while the others will predict the relative logarithmic aleatoric uncertainties (\Sec\ref{sec:methods_aleatoric}). We decide to train our network to predict the Einstein radius ($\theta_E$) and the two components of the complex ellipticity ($\epsilon_x$ and $\epsilon_y$; see e.g. \citealt{Kormann_SIE}) instead of the more common axis ratio and position angle:
\begin{equation}
    \epsilon_x=\frac{1-q^2}{1+q^2}\cos(2\varphi) \quad \epsilon_y=\frac{1-q^2}{1+q^2}\sin(2\varphi)
\end{equation}
As discussed in \citet{Levasseur_modellingBNN} and \citet{Hezaveh_CNN}, this choice increases the algorithm's accuracy. Moreover, as shown in \citet{Levasseur_modellingBNN} and \citet{Kendall_BNN}, we train the algorithm to predict $\log(\sigma^2)$ instead of directly $\sigma$: this choice improves the numerical stability of the algorithm and prevents it from predicting negative variances.

Finally, a reliable confidence interval should consider both the epistemic and aleatoric uncertainties. In this work, we combine the two sources of uncertainty through the procedure previously employed in \citet{Levasseur_modellingBNN}:
\begin{enumerate}
\item For each input image, we predict the mean value ($\mu_i$) and the relative aleatoric uncertainty ($\sigma_i^A$) for the generic parameter $p$ ($\theta_e$,$\epsilon_x$, or $\epsilon_y$);
\item For each parameter $p$, we randomly sample a new value $p_i$ from a Gaussian distribution centred on the predicted value $\mu$ and with a standard deviation equal to $\sigma_i^A$. We assume that the Gaussian form better describes the statistical distribution of the aleatoric uncertainties since these were obtained by minimising a \textit{Gaussian negative log-likelihood} (\Sec\ref{sec:methods_aleatoric});
\item We repeat the first two steps a thousand times for each image. In doing so, we sample the posterior probability of the parameters for each image;
\item Finally, we compute the mean ($\bar{p}$) and standard deviation ($\delta p$) of the distribution given by the $p_i$, obtaining the final value for the considered parameter and its combined uncertainties.
\end{enumerate}

In the following - for the sake of brevity - we will employ the symbols $p$ and $\delta p$ to refer to the mean value of the $p_i$ distribution and the combined uncertainties for each parameter.

\begin{figure*}
\subfloat[][]
{\includegraphics[scale=0.5]{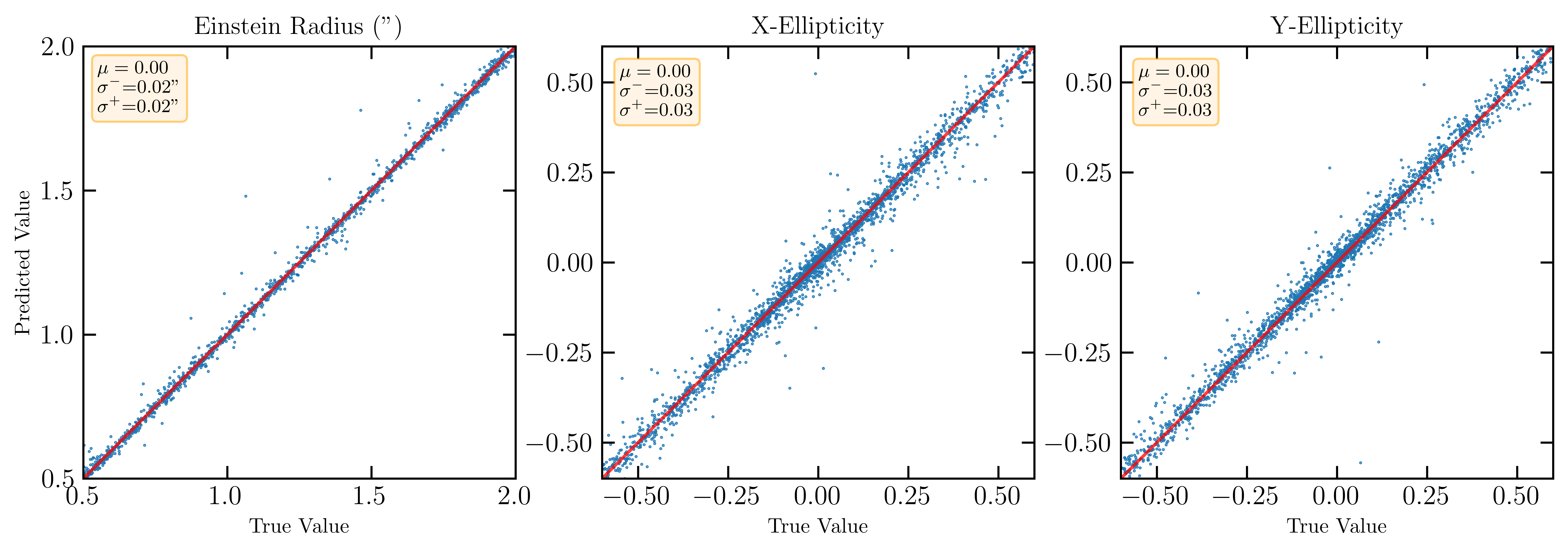}}\\
\subfloat[][]
{\includegraphics[scale=0.5]{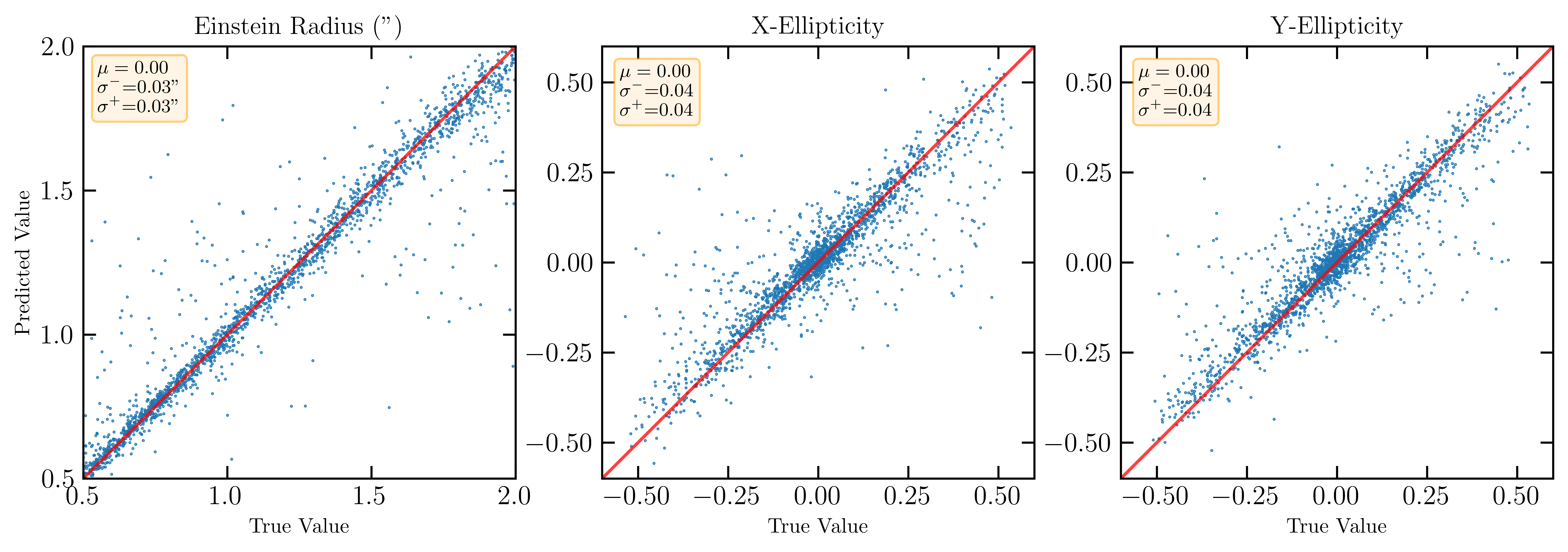}}
\caption{Comparison between the predictions of the \textsc{LeMoN} algorithm and the real values of the SIE parameters for the simulated \textit{Euclid-like} (top row) and \textit{HST-like} (bottom row) datasets. For each plot we also report in the yellow box the \textit{bias} ($\mu$) and the 16th and 84th percentiles of the error distribution ($\sigma^+$ and $\sigma^-$, respectively). Further details in \Sec\ref{sec:results_simulated}.} 
\label{fig:Scatter}
\end{figure*}

\subsection{Training the algorithm}
\label{sec:methods_training}

We perform two different trainings of the \textsc{LeMoN} algorithm: one for each training set discussed in \Sec\ref{sec:data}. We train the algorithm on 75\% of each training set, saving the lasting 25\% for cross-validation. We perform the training with the \textit{stochastic gradient descent} technique, with a batch size of 32 images randomly chosen from the whole dataset. During the training, the algorithm minimises the \textit{Gaussian negative-log-likelihood} introduced in \Eq\ref{eq:loss_ale} through the \textsc{ADAM} optimiser \citep{Kingma_Adam}. We start the training with an initial learning rate of $10^{-3}$. We gradually update the learning rate during the training employing the \textsc{ReduceOnPlateau} callback provided by \textsc{Keras} with a patience parameter concerning the validation loss of five epochs up to $10^{-5}$. An $L_2$ regularisation is also employed to prevent overfitting and to account for the second term present in \Eq\ref{eq:loss}. Finally, during this phase, we employ data augmentation: a common choice in regression and classification problems involving images. It consists of feeding several times the same image to the network, realising a different transformation every time. In this work, each image is translated in the up-down and left-right directions by an integer number of pixels in the range [-4,4]. This step reduces the risk of overfitting since it artificially increases the size of the training set and allows the BNN to learn the translational invariance of the lensing configurations. It is worth underlining that, differently from previous applications \citep[e.g.][]{Petrillo_1,Petrillo_2,Petrillo_3,Gentile_2022}, we do not employ any augmentation involving rotations, flipping, and scaling of the analysed images. This choice is necessary since the predicted parameters are not invariant under these transformations. As discussed in the previous section, dropout layers are employed at both training and testing time, with a \textit{keep-rate} of 0.97. 

The whole trainings require on average 75 epochs and take $\sim2.5h$ each when performed on a single Tesla-K200 GPU freely offered by the cloud-computing platform \textsc{Google Colab}\footnote{\url{https://colab.research.google.com/}}. 

\section{Results}
\label{sec:results}
In this section, we assess the performances of the \textsc{LeMoN} algorithm by applying the BNN to several datasets. After a brief introduction on the metrics employed to evaluate the accuracy of the algorithm and the reliability of the estimated uncertainties, we apply the network to two datasets of simulated images. The results obtained on the \textit{Euclid mocks}, in particular, will be useful to forecast the performances of the \textsc{LeMoN} algorithm when applied to the real data from the forthcoming \textit{Euclid} satellite. These forecasts - however - strongly rely on the hypothesis that the simulations reasonably resemble the characteristics of real lenses. To test this assumption, we also apply the BNN to a dataset of real lenses observed by the HST and compare the results obtained on these systems with those attained on the \textit{HST-like} simulations. Finally, we perform additional tests about the relationship between the accuracy of the algorithm and the employed training sets.

\subsection{Useful metrics}
\label{sec:results_metrics}
Throughout the following sections, we will define an “ideal” lens-modeller as an algorithm able to supply the exact set of SIE parameters for each analysed lens. To compare our results with those expected from such an algorithm, we must introduce some statistical indices (or “metrics” in the following). The choice of these quantities is not unique or trivial and can sensibly affect our description of the results or our ability to compare them with earlier studies present in the current literature. Since our algorithm can supply both a point estimate of the predicted parameters and their confidence intervals, we need more metrics to evaluate these aspects. 

\subsubsection{Overall accuracy}
\label{sec:results_accuracy}
The first property to evaluate is the distance between the mean values predicted by the algorithm and the real ones. In a graphical perspective, the accuracy of a lens-modeller can be assessed through the scatter plots shown in \Fig\ref{fig:Scatter}. The accuracy can also be assessed by measuring the difference $\Delta p$ between each predicted parameter and its real value for each analysed image. Hence, we define the “bias” ($\mu$) as the mean value of the $\Delta p$ distribution computed on the whole dataset and the “standard deviation” ($\sigma$) as the minimum $|\Delta p|$ having a 68\% statistical coverage of the analysed dataset. In addition, we can consider non-symmetrical distributions by computing $\sigma^+$ and $\sigma^-$ as the 16 and 84th percentile of the $\Delta p$ distribution.

\subsubsection{Confidence intervals}
\label{sec:results_uncertaities}
\begin{figure*}
\subfloat[][]
{\includegraphics[width=\columnwidth]{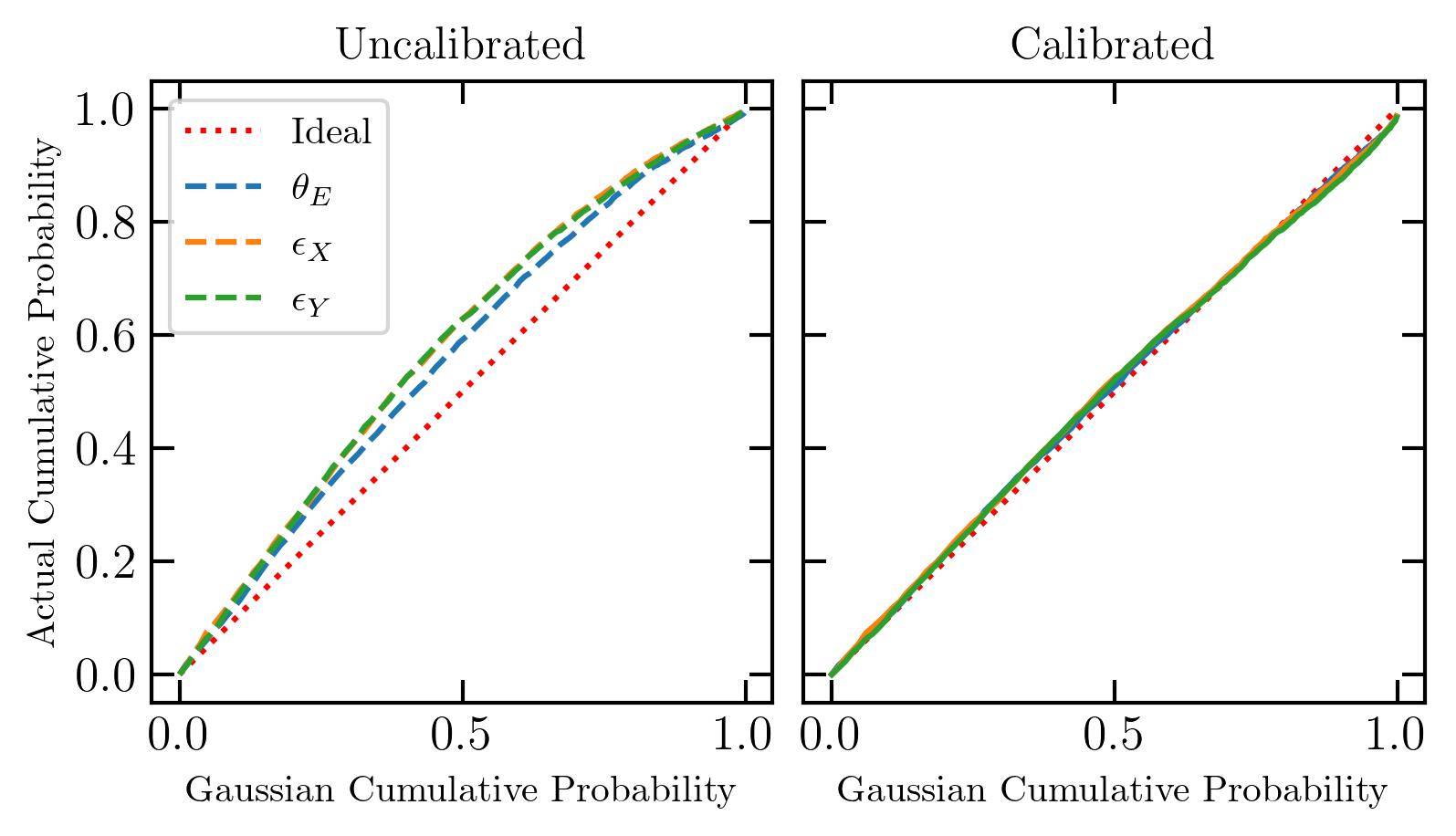}}\qquad
\subfloat[][]
{\includegraphics[width=\columnwidth]{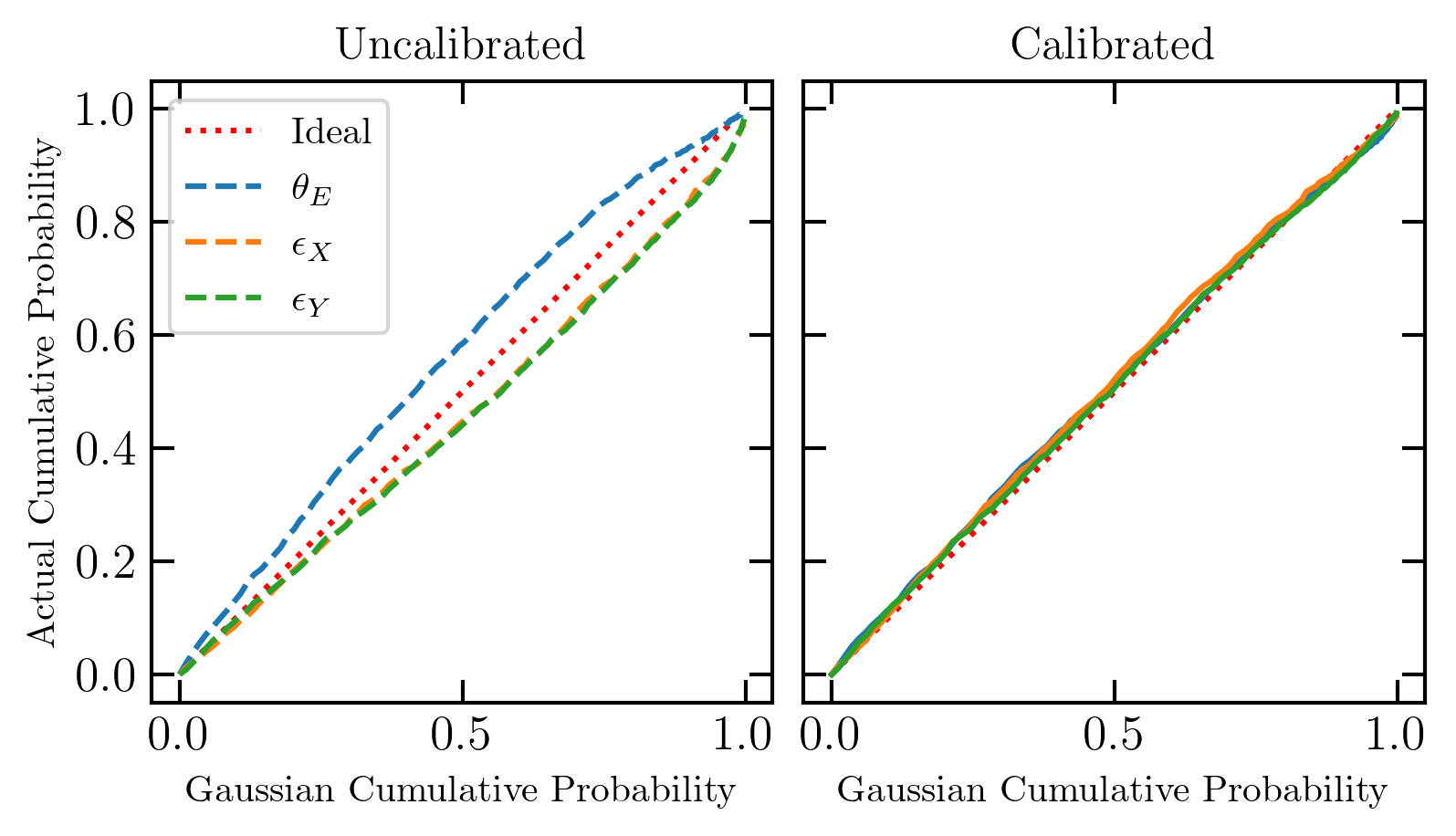}}
\caption{Reliability plots generated for the \textit{HST-like} (a) and for the \textit{Euclid-like} (b) validation sets before and after the calibration procedure described in \Sec\ref{sec:results_calibration}. It is evident how the calibration improves the reliability of the estimated uncertainties and how these are able to reproduce a Gaussian statistical coverage of the analysed datasets. Further details in \Sec\ref{sec:results}.}
\label{fig:Calibration}
\end{figure*}

Evaluating the reliability of the estimated confidence intervals is a more challenging task. Since these values are predicted in an unsupervised framework (see the discussion in \Sec\ref{sec:methods}), we do not have any reference value to compare our results with. In the hypothesis that the uncertainties affecting our algorithm and data can be described with a Gaussian distribution (a reasonable assumption, given the likelihood introduced in \Sec\ref{sec:methods_aleatoric}), we expect that an ideal modeller would provide confidence intervals producing a Gaussian statistical coverage. Therefore, we expect that $\sim68\%$ of the predicted values lie within one estimated $\delta p$ from their actual value, $\sim95\%$ within $2\delta p$ and $\sim$99\% within $3 \delta p$. In most of the earlier studies concerning BNN-based lens-modellers, the reliability of the confidence intervals has been assessed by comparing the statistical coverage at 1, 2 and 3 $\sigma$ computed on the dataset with that expected from a Gaussian distribution. In this work, we employ a more sophisticated method relying on the so-called “\textit{reliability plots}” (\Fig\ref{fig:Calibration}). These graphs are built by computing the cumulative statistical coverage $P(x)$ defined as:
\begin{equation}
    P(x)=\frac{N(|p_i - p_i^\star|<x\delta p_i)}{N_{\rm Tot}}
\end{equation}
where the numerator is the number of systems for which the $\Delta p$ is smaller than $x$ times the related uncertainty $\delta p$. $P(x)$  is evaluated for an ideal Gaussian distribution (on the x-axis) and for the analysed dataset (on the y-axis) for $x$ in the range $[0,5]$. With this definition, an ideally-calibrated algorithm would provide a graph with all the data points lying on the bisector. Similarly, a graph lying in the upper (lower) semi-plane would represent a systematic overestimation (underestimation) of the uncertainties.

\subsection{Calibrating the uncertainties}
\label{sec:results_calibration}

The choice to employ the reliability plots also allows us to perform a more efficient uncertainties calibration with respect to previous studies. BNNs, indeed, are generally not able to provide well-calibrated uncertainties, being affected by a systematic over- or under-estimation of the confidence intervals (see \citealt{Guo_Miscalibration} and references therein for some possible explanations of this phenomenon). Therefore, a successive "calibration step" is often required. This can be performed following different methods \citep[e.g.][]{Levasseur_modellingBNN,Gal_Concrete,Zadrozny_Calibration,Zadrozny_Calibration2}. 

In this work, we employ a slightly modified version of the so-called \textit{platt-scaling} method \citep{Kull_BetaCalibration} employed in \citet{Hortua_CMB}. This method consists in re-scaling the predicted uncertainties by a factor "\textit{s}" ($\sigma \to s\sigma$), in order to minimise the difference between the uncalibrated reliability plot and the ideal one. The procedure is the following:
\begin{enumerate}
    \item We divide the \textit{validation set} in two parts: 30\% for "calibration" and 70\% for "test";
    \item We build the \textit{reliability plot} for the calibration set, employing the original uncertainties estimated by the BNN;
    \item We fit the uncalibrated \textit{reliability plot} with a $\beta$-function defined as
    \begin{equation}
        \beta(x,a,b,c)=\frac{1}{1+e^c\frac{x^a}{(1-x)^b}}
    \end{equation}
    with \textit{a}, \textit{b},\textit{c} $\in  \mathbb{R}$ free parameters fixed during the fitting;
    \item We find the value \textit{s} by minimising\footnote{The minimum is found numerically through the \textsc{Python SciPy} library \citep{Virtanen_SciPy}} the difference between the $\beta$-function and the bisector of the \textit{reliability plot};
    \item We re-scale the uncertainties of the test set predicted by the BNN by the found \textit{s}-factor and build a new "calibrated" reliability plot
\end{enumerate}
\Fig\ref{fig:Calibration} shows the difference between the uncalibrated reliability plot and the calibrated one for the parameters estimated by the \textsc{LeMoN} algorithm. It is clear the improvement in the statistical coverage achieved by this method.

\subsection{Application to simulated data}
\label{sec:results_simulated}

The first assessment of the LeMoN algorithm’s abilities to model vast samples of lenses is performed on two simulated validation sets. We assemble them following the same strategy discussed in \Sec\ref{sec:data}, employing the same division between \textit{HST-} and \textit{Euclid mocks}. To avoid possible biases and perform totally blind cross-validation, we reseed all the random generators used in the procedure: both those employed in the parameters’ sampling and in the background noise generators. Since we are interested in the application to vast datasets, we simulate 10.000 lenses for each dataset. The predictions are performed on the same hardware introduced in \Sec\ref{sec:methods_training} and take $\sim1.5h$ for each dataset, with an average modelling time of less than $0.5$s for each image. The scatter and reliability plots obtained on these datasets are reported in \Figs\ref{fig:Scatter} and \ref{fig:Calibration}.

\begin{figure*}
{\includegraphics[width=\textwidth]{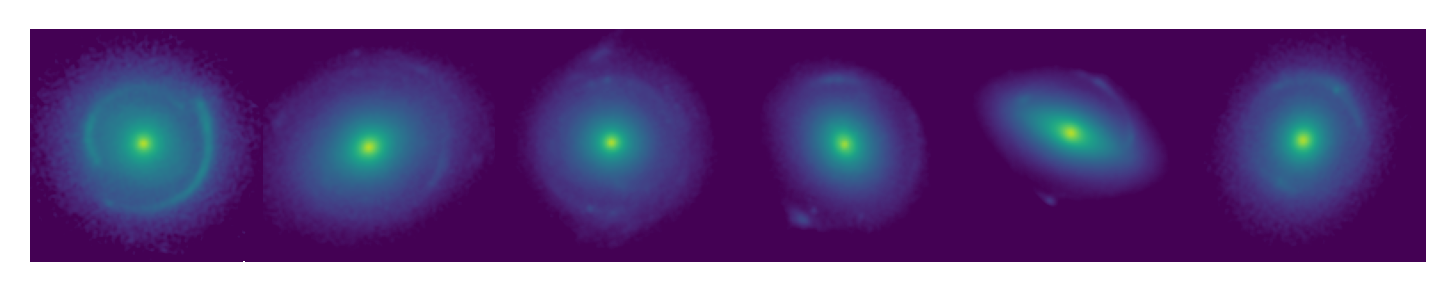}}
\caption{Some examples of the real lenses employed to test the accuracy of the \textsc{LeMoN} algorithm. All the systems were observed by the Hubble Space Telescope during the \textit{Sloan Lens ACS Survey} (SLACS; \citealt{Slacs_I,SLACS_13}). Further details in \Sec\ref{sec:results_real}.}
\label{fig:Lenses_SLACS}
\end{figure*}

\subsection{Further Analysis}
\label{sec:results_further}

The results achieved on these simulated datasets make it interesting to  investigate the impact of the simulations employed to train the algorithm on its accuracy. We further investigate this point by simulating four additional datasets (two with \textit{Euclid-like} properties and two with \textit{HST-like} ones). Two of these datasets are generated following the same procedure described in \Sec\ref{sec:data}, but without adding the light distributions of the deflectors. The importance of these “\textit{arcs-only datasets}” is twofold. On the one hand, these simulated lenses can help us to quantify the loss of accuracy due to the presence of the lens light and the possible improvement in the performances of our algorithm if we employed lens-subtracted systems. On the other hand, the results achieved on these systems can be compared with those obtained by previous studies based on analogous techniques but using lens-subtracted lenses \citep[e.g][]{Hezaveh_CNN,pearson_ModellingCNN} limiting the possible biases due to the different analysed objects.

The other two additional datasets explore the possible impact of the correlation between the light and mass distribution of the deflector. As highlighted by several studies, we do not expect the parameters of the brightness profile of the lens and its matter distribution to be independent. In particular, \citet{Koopmans_06} found an interestingly tight correlation between the parameters of the Sérsic model of the lens and its SIE model:\footnote{Actually, this correlation has been verified for galaxies with $q>0.5$, for more elliptical galaxies some studies found different results \citep[e.g][]{Treu_SWELLS}. Since in this study we are discussing the role of a correlation in itself, we employ the result by \citet{Koopmans_06} for all the lenses in the dataset.}
\begin{equation}
\label{eq:correlations}
\begin{cases}
q_{SIE}=(0.99\pm 0.11)q_L \\
\Delta\theta=\theta_{SIE}-\theta_L=(0\pm3)^\circ
\end{cases}
\end{equation}

Inserting these correlations in the simulated datasets might – in principle – improve the accuracy of our algorithm by augmenting the number of features available to predict the SIE parameters. To further investigate this possibility, we simulate these additional “\textit{correlated-lenses datasets}” through a slightly modified version of the procedure described in \Sec\ref{sec:data_lens}:
\begin{itemize}
    \item \textbf{\textit{Euclid-like} lenses:} The parameters of the Sérsic profile are not randomly sampled from the distribution in \Tab\ref{tab:param}. Once the gravitational arc is simulated, the deflector’s axis ratio and position angle are sampled from a Gaussian distribution centred on the same parameters of the SIE model and with the standard deviations reported in \Eq\ref{eq:correlations} .
    
    \item \textbf{\textit{HST-like} lenses:} The deflector is not randomly chosen from the sample of likely deflectors described in \Sec\ref{sec:data_lens}. Once the gravitational arc is simulated, the $q_L$ and $\varphi_L$ values for the axis ratio and position angle of the deflector, respectively, are sampled from the same Gaussian distribution mentioned above. Therefore, the galaxy with the closest $q$ value in the deflector sample is employed as lens. Analogously, the deflector is rotated to match its original position angle with the new one.
\end{itemize}    
Once these new datasets are simulated (composed of 40.000 images each, as described in \Sec\ref{sec:data}), we perform four additional trainings of the \textsc{LeMoN} algorithm (one for each dataset). Therefore, we apply each network to a corresponding validation set of 10.000 images, simulated following the same prescriptions described before and with a different random seed. The results of these additional cross-validations are reported in \Tab\ref{tab:results}.

\begin{table}
    \centering
    \caption{Mean accuracies achieved by the \textsc{LeMoN} algorithm when applied to the several datasets introduced in \Sec\ref{sec:results_further}. All the values reported in the table are the standard deviation of the error distribution computed as the 68\% statistical coverage of the analysed datasets. Further details are given in \Sec\ref{sec:results_further}.}
    \begin{tabular}{c|cc|cc|cc}
         & \multicolumn{2}{c}{\textit{Standard datasets}}& \multicolumn{2}{c}{\textit{Arcs-only}} & \multicolumn{2}{c}{\textit{"Correlated" lenses}} \\
         \hline
         & \textit{Euclid} & HST & \textit{Euclid} & HST & \textit{Euclid} & HST\\
         & mocks & mocks & mocks & mocks & mocks & mocks \\ 
         \hline
         $\theta_E ('')$ & $0.02$ & $0.03$ & $0.02$ & $0.02$ & $0.02$ & $0.03$ \\
         $\epsilon_X$ & $0.03$ & $0.04$ & $0.02$ & $0.02$ & $0.03$ & $0.04$\\
         $\epsilon_Y$ & $0.03$ & $0.04$ & $0.02$ & $0.02$ & $0.03$ & $0.04$\\
         \hline
    \end{tabular}
    \label{tab:results}
\end{table}

\subsection{Application to real data}
\label{sec:results_real}

\begin{figure*}
\centering
{\includegraphics[scale=0.6]{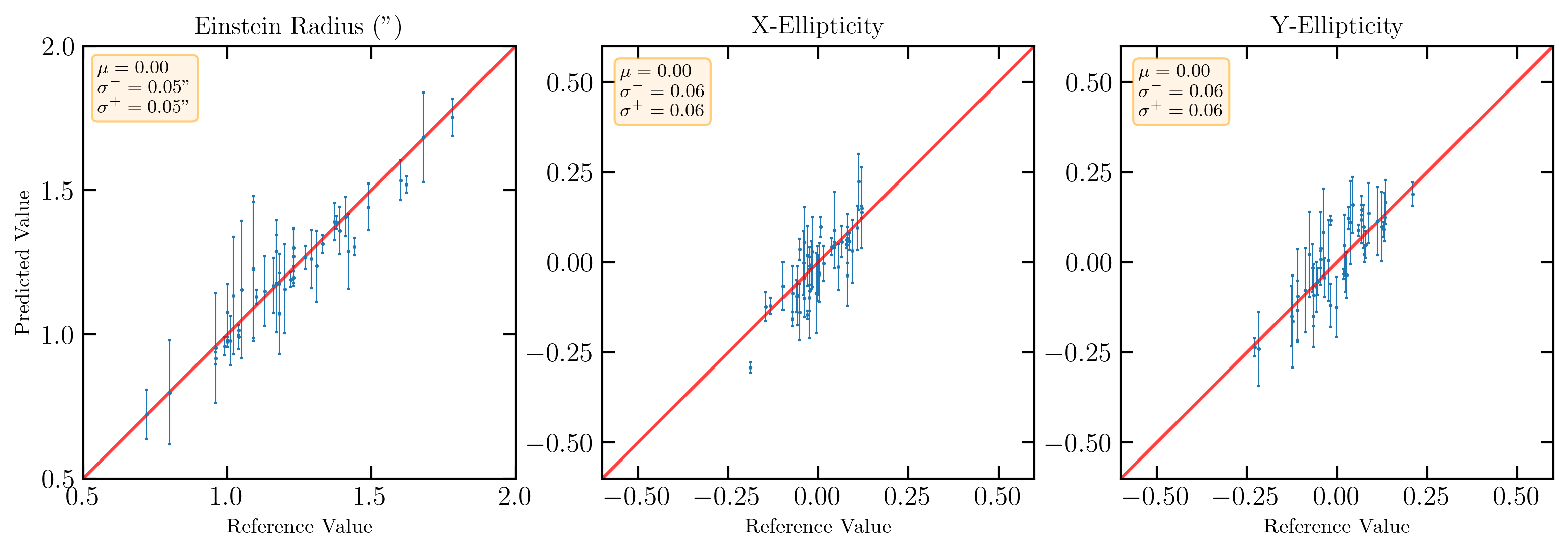}}
\caption{Comparison between the SIE parameters predicted by the \textsc{LeMoN} algorithm and the values obtained by the SLACS collaboration through classical modelling \citep{SLACS_5,SLACS_13}. The uncertainties included in the plot are calibrated as discussed in \Sec\ref{sec:results_calibration}. For each plot we also report in the yellow box the \textit{bias} ($\mu$) and the 16th and 84th percentiles of the error distribution ($\sigma^+$ and $\sigma^-$, respectively). Further details in \Sec\ref{sec:results_real}.}
\label{fig:Scatter_SLACS}
\end{figure*}

\begin{figure}
\centering
{\includegraphics[width=\columnwidth]{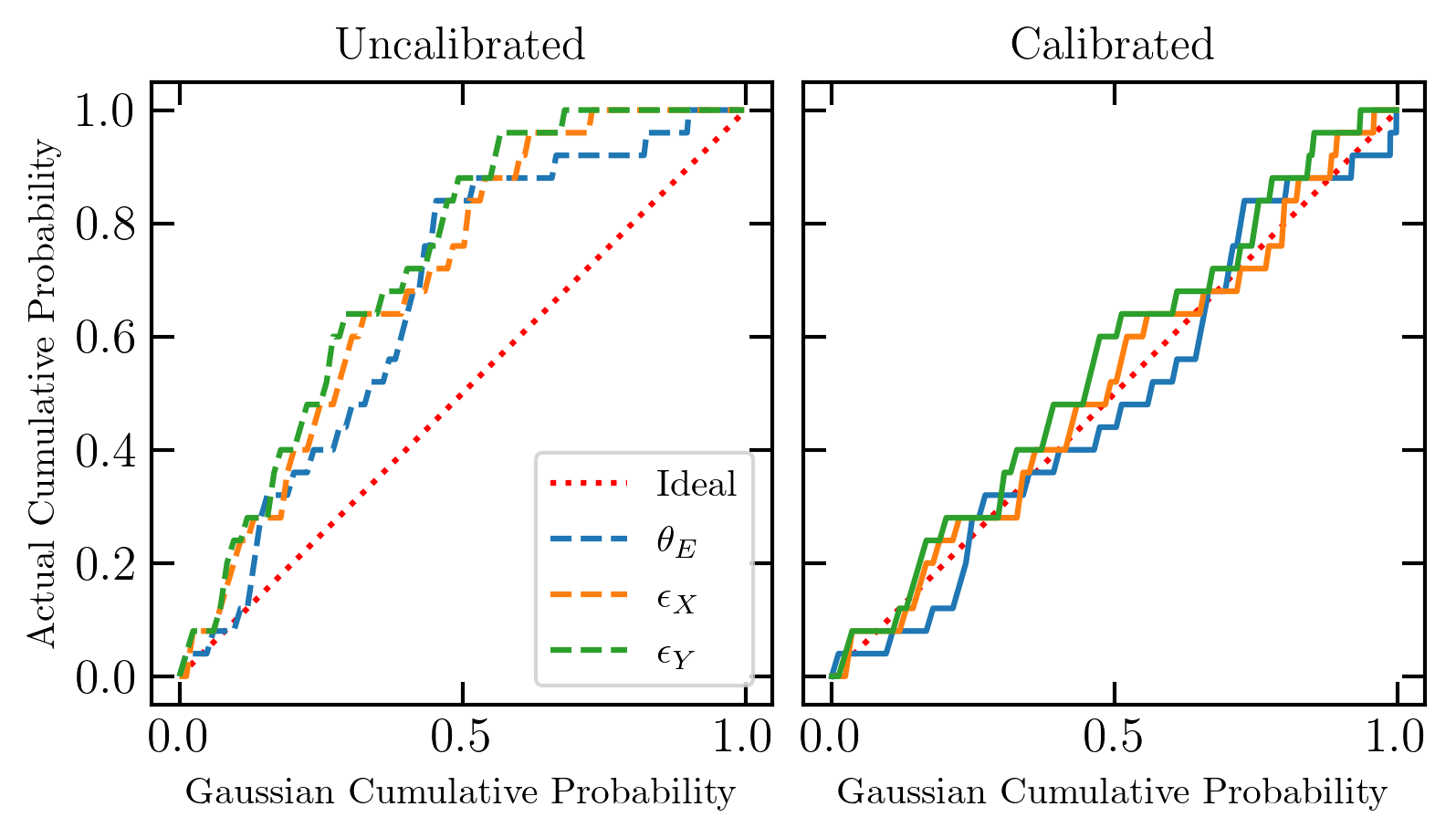}}
\caption{Reliability plots generated by applying the \textsc{LeMoN} algorithm to the pilot sample of real lenses observed by the Hubble Space Telescope in the SLACS survey \citep{Slacs_I,SLACS_13}. The plots are shown before and after the blind-calibration procedure. It is possible to observe the significant increase in the reliability of the estimated uncertainties. Further details are given in \Sec\ref{sec:results_real}.}
\label{fig:Calibration_SLACS}
\end{figure}

To independently assess the accuracy of the \textsc{LeMoN} algorithm in modelling lenses, we apply the BNN trained on the \textit{HST mocks} to a sample of real lenses discovered in the SLACS programme \citep{Slacs_I,SLACS_5,SLACS_13}. Since we do not expect our algorithm to be accurate out of the range of parameters covered by the training set, we only select the systems encompassing these characteristics:
\begin{enumerate}
\item An elliptical central galaxy, as those employed in \Sec\ref{sec:data_lens};
\item The SIE parameters estimated from lens-modelling (as reported in \citealt{SLACS_13} and \citealt{SLACS_5}) within the range summarised in \Tab\ref{tab:param};
\item A SNR integrated on the full arc larger than 5;
\item A brightness ratio between the arc and the central galaxy (i.e. the $\alpha$-factor defined in \Sec\ref{sec:data_lens}) in the range $[0.03,0.2]$.
\end{enumerate}
We retrieve HST imaging in the F814W band for a pilot sample of 42 SLACS lenses having all the above characteristics. \footnote{The HST images are retrieved from the Hubble Legacy Archive: \url{https://hla.stsci.edu/}} (our “\textit{real-lenses sample}” in the following; see some examples in \Fig\ref{fig:Lenses_SLACS}). We extract stamps of these systems with a 131-pixel side, apply a square-root stretch and pass these images to the algorithm. The results obtained by this algorithm application and their comparison with the estimates of the SLACS collaboration are reported in \Figs\ref{fig:Scatter_SLACS} and \ref{fig:Calibration_SLACS}. We underline that the calibration of the uncertainties is performed in a totally-blind way, employing the same “s” factor obtained during the validation step on the simulated \textit{HST mocks} (see \Sec\ref{sec:results_calibration}).

\section{Discussion}
\label{sec:discussion}

This section discusses the main scientific findings obtained through the results described in \Sec\ref{sec:results}.
\begin{itemize}
\item \textbf{Modelling time:} The first noteworthy result achieved by the \textsc{LeMoN} algorithm resides in the time employed to model a vast sample of 10.000 simulated lenses. \citet{pearson_CombinedCBNN} reported – for a blind application of the \textsc{pyAutoLens} lens-modelling algorithm \citep{nightingale_Autolens} – a mean modelling time of 30 minutes, with several outliers requiring up to 2h to be modelled correctly. Our algorithm achieves an average modelling time of 0.5s, independently of the morphology of the analysed system. The improvement of several orders of magnitude in the computational cost of lens modelling between traditional and machine-learning techniques is evident (also thanks to the possibility to train and test the latter algorithms on GPUs, much faster than CPUs). In addition, we underline that the whole procedure employed in this paper to train the algorithm and model the lensed systems is entirely automated, not requiring any human intervention. These results make BNN-based algorithms particularly suitable to model vast datasets of lenses in a computationally efficient way.
 
\item \textbf{Overall Accuracy:} We applied the \textsc{LeMoN} algorithm to two different simulated sets of 10.000 lenses each, retrieving the accuracy summarised in \Sec\ref{sec:results_simulated}. These results are perfectly comparable with other studies present in the current literature and based on similar techniques (CNN or BNN-based lens-modeller, see e.g. \citealt{Hezaveh_CNN,pearson_ModellingCNN,Levasseur_modellingBNN,Schuldt_ModellingCNN}). It is noteworthy to underline how the scatter achieved on the \textit{HST-mocks} is slightly larger than that obtained on the \textit{Euclid-mocks}, albeit the higher resolution of the first simulations. This result is probably due to the more realistic deflector employed in the simulation procedure of the \textit{HST-mocks}. Furthermore, as expected, machine learning algorithms are still not able to outperform or even achieve the same accuracy as classical Bayesian techniques (see \citealt{pearson_CombinedCBNN} for a quantitative comparison between machine-learning and classical methods). Nevertheless, CNN- or BNN-based algorithms can still be employed to provide a first fast modelling of vast samples of lenses to identify the most interesting systems for future follow-up or more accurate modelling with standard techniques. Moreover, as highlighted by \cite{pearson_CombinedCBNN}, machine-learning predictions can be employed as priors for classical Bayesian methods, significantly improving the modelling time with respect to a completely "blind" analysis based on uninformative flat priors.
 
\item \textbf{Calibrated Uncertainties:} As highlighted in \Sec\ref{sec:results_calibration}, the calibration procedure followed to obtain reliable uncertainties is quite effective, producing a Gaussian statistical coverage of the analysed samples (\Fig\ref{fig:Calibration}). The goodness of the procedure is also independently assessed with the blind cross-validation performed in \Secs\ref{sec:results_calibration} and \ref{sec:results_real}. The possibility to predict consistent uncertainties is crucial to allow most of the scientific applications of lensing, where it is fundamental to compare the observational evidence with the theoretical predictions. In addition, high values of the uncertainties can also allow a selection of the contaminants (i.e. the systems for which the modelling procedure failed). Finally, once verified the gaussian statistical coverage, the estimated uncertainties can be employed in classical Bayesian techniques as Gaussian priors, additionally improving the computation time with respect to flat priors.
 
\item \textbf{Different Datasets:} The results achieved on the additional datasets employed in \Sec\ref{sec:results_further} translate into several findings concerning the relationship between the employed training sets and the algorithm's accuracy. Firstly, as expected, the subtraction of the lens light increases the accuracy of the modelling (up to 100\%, see \Tab\ref{tab:results}). We underline, however, that the subtraction of the lens light is problematic for two main reasons. Firstly, this process generally relies on parametric modelling of the deflector: wrong modelling can produce a severe bias in the deblending procedure and the inferred shape of the gravitational arc. Secondly, this procedure is not completely automatic but often requires a non-trivial human intervention to select the lensing features. The latter reason, in particular, strongly affects the possibility of modelling vast samples of lenses automatically. It is also worth noting that the improvement in the accuracy due to the better image quality achieved by HST is less significant in the "\textit{arcs-only}" dataset.

A second – more counter-intuitive – finding consists in the identical accuracy obtained by the \textsc{LeMoN} algorithm when applied to the correlated and uncorrelated lens datasets. A possible explanation of this result could reside in the ability of our algorithm to well discriminate between the gravitational arc and the lensing galaxy, basing its prediction only on the feature belonging to the first one and ignoring the shape and orientation of the latter.
 
\item \textbf{Real Lenses:} When applied to the set of real lenses from the SLACS sample (\Sec\ref{sec:results_real}), the algorithm achieves the results summarised in \Figs\ref{fig:Scatter_SLACS} and \ref{fig:Calibration_SLACS}.  We underline that - even though the bias is always consistent with zero - the algorithm produces a higher scatter between predicted and real values when passing from simulated to real data. This phenomenon is not new, as discussed in several studies concerning lens-finding with machine-learning algorithms \citep[e.g.][]{Petrillo_1,Petrillo_2,Petrillo_3,Gentile_2022}. Our interpretation of this result resides in the still-not-sufficient realism of our simulations. In the near future, it is possible to think of a training set composed only of real lenses observed in a wide survey such as those performed with \textit{Euclid} or the \textit{Vera Rubin Observatory}. At the moment, however, the only alternative consists in generating more reliable simulations. A possible improvement could be the employment of real galaxies as lensed sources \citep[e.g.][]{pearson_CombinedCBNN} or the increasing of the complexity of the deflector by employing more complex lens models (e.g. adding an external shear component, e.g. \citealt{Keeton_Sie}), or a better accounting for the matter distribution on the line-of-sight of the lens \citep[e.g.][]{Fleury_21} and likely perturbation to the lensing potential \citep[e.g.][]{Vernardos_22,Galan_22}. These improvements will be addressed in detail in the forthcoming papers of the \textsc{LeMoN} series.
\end{itemize}

\subsection{Comparison with the literature}
\label{sec:comparison}

The unprecedented number of lenses expected by next-generation facilities will represent a significant challenge to our analysis capabilities. The urgency of this issue justifies the presence, in the current literature, of at least two groups of studies aiming to automatise the lens-modelling procedure. The first of these is focused on existing “classical” algorithms, trying to reduce their computational cost \citep[e.g.][]{Gu_Gigalens} or to decrease the needed human intervention \citep[e.g.][]{Etherington_autolens}. A second group (also including this paper) tries to change the paradigm by involving machine learning algorithms in the task. 
This section focuses on comparing our code \textsc{LeMoN} with the algorithms proposed by this second group. However, it is crucial to underline how a fair comparison between the different codes should consider the difference between the datasets employed to train and test the various algorithms. For instance, some studies \citep[][Fagin et al. in prep]{Hezaveh_CNN,Levasseur_modellingBNN} focused on HST- or Euclid- like simulations, while others used ground-based simulated images \citep{bom_2019,Schuldt_ModellingCNN}. It is evident how the better image quality obtained by space-based facilities allows a more accurate lens-modelling (see, e.g. the result by \citealt{pearson_ModellingCNN} employing both kinds of simulations). Similar effects are expected by the employment of more realistic simulations. Using real galaxies as deflectors and lensed sources \citep{Hezaveh_CNN,pearson_CombinedCBNN,Schuldt_ModellingCNN} increases the realism of the simulations but also decreases the accuracy of the lens-modelling, as shown by \citet{pearson_CombinedCBNN}. Analogously, the subtraction of the lens light generally produces higher accuracies with respect to studies employing the complete images \citep[see e.g.][]{Hezaveh_CNN,pearson_ModellingCNN}. Since all these issues could be overcome only by training and testing all the codes on the same dataset, we will neglect the likely effects due to the datasets to focus on other important differences.
\begin{itemize}
\item The first of these concerns the kind of algorithm employed. \citet{Hezaveh_CNN, pearson_ModellingCNN,pearson_CombinedCBNN} and \citet{Schuldt_ModellingCNN} used CNN-based lens-modeller. The most significant difference between these codes and \textsc{LeMoN} is the possibility for our algorithm to estimate accurate uncertainties for the predicted parameters. Furthermore, the broad consistency of the accuracies reported by the different codes suggests that the estimation of the uncertainties does not sensibly affects the accuracy of the algorithm’s point estimates.
\item For the algorithms based on a BNN \citep{Levasseur_modellingBNN,bom_2019,Schuldt_22}, one of the main differences is the calibration performed on the uncertainties predicted by the network. While the other codes changed the keep-rate of the dropout layers to achieve a Gaussian coverage of the test-set, the \textsc{LeMoN} algorithm is the only one to employ the \textit{a-posteriori} procedure discussed in \Sec\ref{sec:results_calibration}. As shown in \Figs\ref{fig:Calibration} and \ref{fig:Calibration_SLACS}, our procedure allows an effective calibration of the uncertainties both on simulated and real data.
\item A further consideration concerns the kind of architecture employed by the different codes. While \citet{pearson_ModellingCNN} and \citet{Schuldt_ModellingCNN} used a “classic” CNN, other studies employed a ResNet (i.e. the same architecture of \textsc{LeMoN}, but with different amounts of layers) and others much more complex architectures \citep{Hezaveh_CNN,Levasseur_modellingBNN,bom_2019,Schuldt_22}. These differences strongly impact the dimension of the datasets required to successfully train the algorithms and – therefore – the possibility of training them only by employing the first data provided by Euclid or LSST. The slight differences in the accuracies achieved by these codes could suggest a minor role of the employed architecture in the effectiveness of the lens modelling. 
\item Finally, it is noteworthy to underline how – among the aforementioned studies – only \citet{Hezaveh_CNN} validated the accuracies achieved on the simulations with an application to a pilot sample of 9 real lenses observed in the SL2S program \citep{Sonnenfeld_SL2S}. As discussed in \Sec\ref{sec:results_real}, this procedure can highlight a significant difference between the accuracies, giving precious insights about the realism of the simulations employed in the training and test phases.
\end{itemize}

\section{Conclusion}
\label{sec:conclusions}

In this work, we presented \textsc{LeMoN}: a new machine learning algorithm able to perform the fast automated analysis of strong gravitational lenses. The algorithm is based on a \textit{Bayesian Neural Network}: a new generation of machine learning algorithms able to associate reliable confidence intervals to the predicted parameters. This paper is the first of the \textsc{LeMoN} series and is focused on the training of the algorithm and its first blind application to several datasets of both simulated and a pilot sample of real gravitational lenses observed with Hubble Space Telescope (HST) as a part of the SLACS program \citep{Slacs_I,SLACS_13}. \textsc{LeMoN} has been trained on two simulated datasets generated to resemble the imaging capabilities of the HST and the forthcoming ESA \textit{Euclid} satellite. The main task of the algorithm is to estimate the three parameters of the Singular Isothermal Ellipsoid (SIE) model ($\theta_E$, $\epsilon_x$ and $\epsilon_y$) and the related uncertainties. After the training, the algorithm has been applied to two simulated datasets generated following the same simulation procedure, to independently assess the accuracy of the estimates and the reliability of the modelled uncertainties. 

The main scientific result consists in the modelling time of just 0.5s for each image, at least two orders of magnitudes lower than classical techniques. Moreover, the results obtained on these datasets are perfectly comparable with the previous studies based on analogous machine-learning algorithms and present in the current literature. Finally, since the estimated uncertainties are able to produce a Gaussian statistical coverage of the analysed dataset, we concluded that the confidence intervals are perfectly reliable and able to allow a vast set of scientific application. To provide an additional confirmation of the efficiency and accuracy of our algorithm, we applied the \textsc{LeMoN} network to a pilot sample of real lenses discovered in the SLACS survey. We found that the algorithm is able to predict the parameters of each lens without any significant bias and to provide well-calibrated uncertainties. However, the application to real data provided a higher scatter between the real and predicted values when compared to the results obtained on the simulated datasets. This issue – most likely explainable with the not-sufficient realism of the simulations in the training set – will be afforded in detail in the forthcoming papers of the \textsc{LeMoN} series. 

\section*{Acknowledgements}
FG acknowledges the support from grant PRIN MIUR 2017-20173ML3WW\_001. \\
FG and CT thank Georgios Vernardos and Jose Maria Diego for the fruitful discussion on the paper.\\
This research is based on observations made with the NASA/ESA Hubble Space Telescope obtained from the Space Telescope Science Institute, which is operated by the Association of Universities for Research in Astronomy, Inc., under NASA contract NAS 5–26555. These observations are associated with programs 10563, 10494, 10886, 10798.

\section*{Data Availability}
The data that support the findings of this study are available from the corresponding author, FG, upon reasonable request. The code employed in this study is freely available in a GitHub repository: \url{https://github.com/fab-gentile/LeMoN}

\bibliographystyle{mnras}
\bibliography{Paper_LeMoN}

\begin{thebibliography}{}
\makeatletter
\relax
\def\mn@urlcharsother{\let\do\@makeother \do\$\do\&\do\#\do\^\do\_\do\%\do\~}
\def\mn@doi{\begingroup\mn@urlcharsother \@ifnextchar [ {\mn@doi@}
  {\mn@doi@[]}}
\def\mn@doi@[#1]#2{\def\@tempa{#1}\ifx\@tempa\@empty \href
  {http://dx.doi.org/#2} {doi:#2}\else \href {http://dx.doi.org/#2} {#1}\fi
  \endgroup}
\def\mn@eprint#1#2{\mn@eprint@#1:#2::\@nil}
\def\mn@eprint@arXiv#1{\href {http://arxiv.org/abs/#1} {{\tt arXiv:#1}}}
\def\mn@eprint@dblp#1{\href {http://dblp.uni-trier.de/rec/bibtex/#1.xml}
  {dblp:#1}}
\def\mn@eprint@#1:#2:#3:#4\@nil{\def\@tempa {#1}\def\@tempb {#2}\def\@tempc
  {#3}\ifx \@tempc \@empty \let \@tempc \@tempb \let \@tempb \@tempa \fi \ifx
  \@tempb \@empty \def\@tempb {arXiv}\fi \@ifundefined
  {mn@eprint@\@tempb}{\@tempb:\@tempc}{\expandafter \expandafter \csname
  mn@eprint@\@tempb\endcsname \expandafter{\@tempc}}}

\bibitem[\protect\citeauthoryear{Abadi et~al.,}{Abadi
  et~al.}{2015}]{Abadi_TensorFlow}
Abadi M.,  et~al., 2015, {TensorFlow}: Large-Scale Machine Learning on
  Heterogeneous Systems, \url {https://www.tensorflow.org/}

\bibitem[\protect\citeauthoryear{{Aniyan} \& {Thorat}}{{Aniyan} \&
  {Thorat}}{2017}]{Aniyan_Morphology}
{Aniyan} A.~K.,  {Thorat} K.,  2017, \mn@doi [\apjs]
  {10.3847/1538-4365/aa7333}, \href
  {https://ui.adsabs.harvard.edu/abs/2017ApJS..230...20A} {230, 20}

\bibitem[\protect\citeauthoryear{Auger, Treu, Gavazzi, Bolton, Koopmans  \&
  Marshall}{Auger et~al.}{2010a}]{Auger_IMF}
Auger M.~W.,  Treu T.,  Gavazzi R.,  Bolton A.~S.,  Koopmans L. V.~E.,
  Marshall P.~J.,  2010a, \mn@doi [The Astrophysical Journal Letters]
  {10.1088/2041-8205/721/2/L163}, 721, L163

\bibitem[\protect\citeauthoryear{Auger, Treu, Bolton, Gavazzi, Koopmans,
  Marshall, Moustakas  \& Burles}{Auger et~al.}{2010b}]{Auger_DM}
Auger M.~W.,  Treu T.,  Bolton A.~S.,  Gavazzi R.,  Koopmans L. V.~E.,
  Marshall P.~J.,  Moustakas L.~A.,   Burles S.,  2010b, \mn@doi [The
  Astrophysical Journal] {10.1088/0004-637X/724/1/511}, 724, 511

\bibitem[\protect\citeauthoryear{Barnab{\`e}, Spiniello, Koopmans, Trager,
  Czoske  \& Treu}{Barnab{\`e} et~al.}{2013}]{barnabe_IMF}
Barnab{\`e} M.,  Spiniello C.,  Koopmans L. V.~E.,  Trager S.~C.,  Czoske O.,
  Treu T.,  2013, \mn@doi [Monthly Notices of the Royal Astronomical Society]
  {10.1093/mnras/stt1727}, 436, 253

\bibitem[\protect\citeauthoryear{Bartelmann}{Bartelmann}{2010}]{bartelmann_theory}
Bartelmann M.,  2010, \mn@doi [Classical and Quantum Gravity]
  {10.1088/0264-9381/27/23/233001}, 27, 233001

\bibitem[\protect\citeauthoryear{{Birrer} \& {Amara}}{{Birrer} \&
  {Amara}}{2018}]{Birrer_2018}
{Birrer} S.,  {Amara} A.,  2018, \mn@doi [Physics of the Dark Universe]
  {10.1016/j.dark.2018.11.002}, \href
  {https://ui.adsabs.harvard.edu/abs/2018PDU....22..189B} {22, 189}

\bibitem[\protect\citeauthoryear{{Bolton}, {Burles}, {Koopmans}, {Treu}  \&
  {Moustakas}}{{Bolton} et~al.}{2006}]{Slacs_I}
{Bolton} A.~S.,  {Burles} S.,  {Koopmans} L. V.~E.,  {Treu} T.,   {Moustakas}
  L.~A.,  2006, \mn@doi [\apj] {10.1086/498884}, \href
  {https://ui.adsabs.harvard.edu/abs/2006ApJ...638..703B} {638, 703}

\bibitem[\protect\citeauthoryear{{Bolton}, {Burles}, {Koopmans}, {Treu},
  {Gavazzi}, {Moustakas}, {Wayth}  \& {Schlegel}}{{Bolton}
  et~al.}{2008}]{SLACS_5}
{Bolton} A.~S.,  {Burles} S.,  {Koopmans} L. V.~E.,  {Treu} T.,  {Gavazzi} R.,
  {Moustakas} L.~A.,  {Wayth} R.,   {Schlegel} D.~J.,  2008, \mn@doi [\apj]
  {10.1086/589327}, \href
  {https://ui.adsabs.harvard.edu/abs/2008ApJ...682..964B} {682, 964}

\bibitem[\protect\citeauthoryear{Bom, Poh, Nord, {Blanco-Valentin}  \&
  Dias}{Bom et~al.}{2019}]{bom_2019}
Bom C.,  Poh J.,  Nord B.,  {Blanco-Valentin} M.,   Dias L.,  2019, arXiv
  e-prints, 1911, arXiv:1911.06341

\bibitem[\protect\citeauthoryear{Canameras et~al.,}{Canameras
  et~al.}{2020}]{Canameras_HoliSmokes}
Canameras R.,  et~al., 2020, arXiv e-prints, 2004, arXiv:2004.13048

\bibitem[\protect\citeauthoryear{Charnock, {Perreault-Levasseur}  \&
  Lanusse}{Charnock et~al.}{2020}]{Charnock_BNN}
Charnock T.,  {Perreault-Levasseur} L.,   Lanusse F.,  2020, arXiv e-prints,
  2006, arXiv:2006.01490

\bibitem[\protect\citeauthoryear{Chatterjee}{Chatterjee}{2019}]{Chatterjee_PhD}
Chatterjee S.,  2019, PhD thesis, University of Groningen / University of
  Groningen

\bibitem[\protect\citeauthoryear{Chollet et~al.}{Chollet
  et~al.}{2015}]{Chollet_keras}
Chollet F.,  et~al., 2015, Keras, \url{https://keras.io}

\bibitem[\protect\citeauthoryear{{Cobb} et~al.,}{{Cobb}
  et~al.}{2019}]{Cobb_2019}
{Cobb} A.~D.,  et~al., 2019, \mn@doi [\aj] {10.3847/1538-3881/ab2390}, \href
  {https://ui.adsabs.harvard.edu/abs/2019AJ....158...33C} {158, 33}

\bibitem[\protect\citeauthoryear{Collett}{Collett}{2015}]{Collett_LensPop}
Collett T.~E.,  2015, \mn@doi [The Astrophysical Journal]
  {10.1088/0004-637X/811/1/20}, \href
  {https://ui.adsabs.harvard.edu/abs/2015ApJ...811...20C} {811, 20}

\bibitem[\protect\citeauthoryear{{Congdon} \& {Keeton}}{{Congdon} \&
  {Keeton}}{2018}]{Congdon_Lensing}
{Congdon} A.~B.,  {Keeton} C.~R.,  2018, {Principles of Gravitational Lensing:
  Light Deflection as a Probe of Astrophysics and Cosmology},
  \mn@doi{10.1007/978-3-030-02122-1.
}

\bibitem[\protect\citeauthoryear{{Covone} et~al.,}{{Covone}
  et~al.}{2009}]{Covone_2009}
{Covone} G.,  et~al., 2009, \mn@doi [\apj] {10.1088/0004-637X/691/1/531}, \href
  {https://ui.adsabs.harvard.edu/abs/2009ApJ...691..531C} {691, 531}

\bibitem[\protect\citeauthoryear{{Cropper} et~al.,}{{Cropper}
  et~al.}{2018}]{Cropper_VIS}
{Cropper} M.,  et~al., 2018, in {Lystrup} M.,  {MacEwen} H.~A.,  {Fazio} G.~G.,
   {Batalha} N.,  {Siegler} N.,   {Tong} E.~C.,  eds,  Society of Photo-Optical
  Instrumentation Engineers (SPIE) Conference Series Vol. 10698, Space
  Telescopes and Instrumentation 2018: Optical, Infrared, and Millimeter Wave.
  p. 1069828, \mn@doi{10.1117/12.2315372}

\bibitem[\protect\citeauthoryear{Dalal \& Kochanek}{Dalal \&
  Kochanek}{2002}]{dalal_substructure}
Dalal N.,  Kochanek C.~S.,  2002, \mn@doi [The Astrophysical Journal]
  {10.1086/340303}, 572, 25

\bibitem[\protect\citeauthoryear{{Dieleman}, {Willett}  \& {Dambre}}{{Dieleman}
  et~al.}{2015}]{Dielman_Morphology}
{Dieleman} S.,  {Willett} K.~W.,   {Dambre} J.,  2015, \mn@doi [\mnras]
  {10.1093/mnras/stv632}, \href
  {https://ui.adsabs.harvard.edu/abs/2015MNRAS.450.1441D} {450, 1441}

\bibitem[\protect\citeauthoryear{{Dom{\'\i}nguez S{\'a}nchez},
  {Huertas-Company}, {Bernardi}, {Tuccillo}  \& {Fischer}}{{Dom{\'\i}nguez
  S{\'a}nchez} et~al.}{2018}]{Dominguez_Morphology}
{Dom{\'\i}nguez S{\'a}nchez} H.,  {Huertas-Company} M.,  {Bernardi} M.,
  {Tuccillo} D.,   {Fischer} J.~L.,  2018, \mn@doi [\mnras]
  {10.1093/mnras/sty338}, \href
  {https://ui.adsabs.harvard.edu/abs/2018MNRAS.476.3661D} {476, 3661}

\bibitem[\protect\citeauthoryear{Einstein}{Einstein}{1915}]{einstein_1915}
Einstein A.,  1915, Sitzungsberichte der K\"oniglich Preu\ss ischen Akademie
  der Wissenschaften (Berlin), Seite 831-839.

\bibitem[\protect\citeauthoryear{{Eisenstein} et~al.,}{{Eisenstein}
  et~al.}{2001}]{Eisenstein_LRG}
{Eisenstein} D.~J.,  et~al., 2001, \mn@doi [\aj] {10.1086/323717}, \href
  {https://ui.adsabs.harvard.edu/abs/2001AJ....122.2267E} {122, 2267}

\bibitem[\protect\citeauthoryear{{Escamilla-Rivera}, {Carvajal Quintero}  \&
  {Capozziello}}{{Escamilla-Rivera} et~al.}{2020}]{Rivera_2020}
{Escamilla-Rivera} C.,  {Carvajal Quintero} M.~A.,   {Capozziello} S.,  2020,
  \mn@doi [\jcap] {10.1088/1475-7516/2020/03/008}, \href
  {https://ui.adsabs.harvard.edu/abs/2020JCAP...03..008E} {2020, 008}

\bibitem[\protect\citeauthoryear{{Etherington} et~al.,}{{Etherington}
  et~al.}{2022}]{Etherington_autolens}
{Etherington} A.,  et~al., 2022, arXiv e-prints, \href
  {https://ui.adsabs.harvard.edu/abs/2022arXiv220209201E} {p. arXiv:2202.09201}

\bibitem[\protect\citeauthoryear{{Fleury}, {Larena}  \& {Uzan}}{{Fleury}
  et~al.}{2021}]{Fleury_21}
{Fleury} P.,  {Larena} J.,   {Uzan} J.-P.,  2021, \mn@doi [\jcap]
  {10.1088/1475-7516/2021/08/024}, \href
  {https://ui.adsabs.harvard.edu/abs/2021JCAP...08..024F} {2021, 024}

\bibitem[\protect\citeauthoryear{{Fluri}, {Kacprzak}, {Lucchi}, {Refregier},
  {Amara}, {Hofmann}  \& {Schneider}}{{Fluri} et~al.}{2019}]{Fluri_Cosmology}
{Fluri} J.,  {Kacprzak} T.,  {Lucchi} A.,  {Refregier} A.,  {Amara} A.,
  {Hofmann} T.,   {Schneider} A.,  2019, \mn@doi [\prd]
  {10.1103/PhysRevD.100.063514}, \href
  {https://ui.adsabs.harvard.edu/abs/2019PhRvD.100f3514F} {100, 063514}

\bibitem[\protect\citeauthoryear{{Gal} \& {Ghahramani}}{{Gal} \&
  {Ghahramani}}{2015a}]{Gal_2016}
{Gal} Y.,  {Ghahramani} Z.,  2015a, arXiv e-prints, \href
  {https://ui.adsabs.harvard.edu/abs/2015arXiv150602142G} {p. arXiv:1506.02142}

\bibitem[\protect\citeauthoryear{{Gal} \& {Ghahramani}}{{Gal} \&
  {Ghahramani}}{2015b}]{Gal_2015}
{Gal} Y.,  {Ghahramani} Z.,  2015b, arXiv e-prints, \href
  {https://ui.adsabs.harvard.edu/abs/2015arXiv150602158G} {p. arXiv:1506.02158}

\bibitem[\protect\citeauthoryear{{Gal}, {Hron}  \& {Kendall}}{{Gal}
  et~al.}{2017}]{Gal_Concrete}
{Gal} Y.,  {Hron} J.,   {Kendall} A.,  2017, arXiv e-prints, \href
  {https://ui.adsabs.harvard.edu/abs/2017arXiv170507832G} {p. arXiv:1705.07832}

\bibitem[\protect\citeauthoryear{{Galan}, {Vernardos}, {Peel}, {Courbin}  \&
  {Starck}}{{Galan} et~al.}{2022}]{Galan_22}
{Galan} A.,  {Vernardos} G.,  {Peel} A.,  {Courbin} F.,   {Starck} J.-L.,
  2022, arXiv e-prints, \href
  {https://ui.adsabs.harvard.edu/abs/2022arXiv220705763G} {p. arXiv:2207.05763}

\bibitem[\protect\citeauthoryear{{Gavazzi}, {Treu}, {Marshall}, {Brault}  \&
  {Ruff}}{{Gavazzi} et~al.}{2012}]{SL2S_first}
{Gavazzi} R.,  {Treu} T.,  {Marshall} P.~J.,  {Brault} F.,   {Ruff} A.,  2012,
  \mn@doi [\apj] {10.1088/0004-637X/761/2/170}, \href
  {https://ui.adsabs.harvard.edu/abs/2012ApJ...761..170G} {761, 170}

\bibitem[\protect\citeauthoryear{{Gentile} et~al.,}{{Gentile}
  et~al.}{2022}]{Gentile_2022}
{Gentile} F.,  et~al., 2022, \mn@doi [\mnras] {10.1093/mnras/stab3386}, \href
  {https://ui.adsabs.harvard.edu/abs/2022MNRAS.510..500G} {510, 500}

\bibitem[\protect\citeauthoryear{Goodfellow, Bengio  \& Courville}{Goodfellow
  et~al.}{2016}]{Goodfellow_ML}
Goodfellow I.,  Bengio Y.,   Courville A.,  2016, Deep Learning.
{MIT Press}

\bibitem[\protect\citeauthoryear{{Gu} et~al.,}{{Gu} et~al.}{2022}]{Gu_Gigalens}
{Gu} A.,  et~al., 2022, \mn@doi [\apj] {10.3847/1538-4357/ac6de4}, \href
  {https://ui.adsabs.harvard.edu/abs/2022ApJ...935...49G} {935, 49}

\bibitem[\protect\citeauthoryear{Guo, Pleiss, Sun  \& Weinberger}{Guo
  et~al.}{2017}]{Guo_Miscalibration}
Guo C.,  Pleiss G.,  Sun Y.,   Weinberger K.~Q.,  2017, in Proceedings of the
  34th International Conference on Machine Learning - Volume 70. ICML'17.
p. 1321–1330

\bibitem[\protect\citeauthoryear{{He}, {Zhang}, {Ren}  \& {Sun}}{{He}
  et~al.}{2015}]{He_ResNet}
{He} K.,  {Zhang} X.,  {Ren} S.,   {Sun} J.,  2015, arXiv e-prints, \href
  {https://ui.adsabs.harvard.edu/abs/2015arXiv151203385H} {p. arXiv:1512.03385}

\bibitem[\protect\citeauthoryear{He et~al.,}{He et~al.}{2020}]{He_KiDS}
He Z.,  et~al., 2020, \mn@doi [Monthly Notices of the Royal Astronomical
  Society] {10.1093/mnras/staa1917}, 497, 556

\bibitem[\protect\citeauthoryear{Hezaveh, Dalal, Holder, Kisner, Kuhlen  \&
  Levasseur}{Hezaveh et~al.}{2016}]{Hezaveh_16}
Hezaveh Y.,  Dalal N.,  Holder G.,  Kisner T.,  Kuhlen M.,   Levasseur L.~P.,
  2016, \mn@doi [Journal of Cosmology and Astroparticle Physics]
  {10.1088/1475-7516/2016/11/048}, 2016, 048

\bibitem[\protect\citeauthoryear{Hezaveh, Perreault~Levasseur  \&
  Marshall}{Hezaveh et~al.}{2017}]{Hezaveh_CNN}
Hezaveh Y.~D.,  Perreault~Levasseur L.,   Marshall P.~J.,  2017, \mn@doi
  [Nature] {10.1038/nature23463}, 548, 555

\bibitem[\protect\citeauthoryear{{Hort{\'u}a}, {Volpi}, {Marinelli}  \&
  {Malag{\`o}}}{{Hort{\'u}a} et~al.}{2020}]{Hortua_CMB}
{Hort{\'u}a} H.~J.,  {Volpi} R.,  {Marinelli} D.,   {Malag{\`o}} L.,  2020,
  \mn@doi [\prd] {10.1103/PhysRevD.102.103509}, \href
  {https://ui.adsabs.harvard.edu/abs/2020PhRvD.102j3509H} {102, 103509}

\bibitem[\protect\citeauthoryear{{Huchra}, {Gorenstein}, {Kent}, {Shapiro},
  {Smith}, {Horine}  \& {Perley}}{{Huchra} et~al.}{1985}]{Huchra_1985}
{Huchra} J.,  {Gorenstein} M.,  {Kent} S.,  {Shapiro} I.,  {Smith} G.,
  {Horine} E.,   {Perley} R.,  1985, \mn@doi [\aj] {10.1086/113777}, \href
  {https://ui.adsabs.harvard.edu/abs/1985AJ.....90..691H} {90, 691}

\bibitem[\protect\citeauthoryear{Jacobs et~al.,}{Jacobs
  et~al.}{2019a}]{Jacobs_DES}
Jacobs C.,  et~al., 2019a, \mn@doi [The Astrophysical Journal Supplement
  Series] {10.3847/1538-4365/ab26b6}, 243, 17

\bibitem[\protect\citeauthoryear{Jacobs et~al.,}{Jacobs
  et~al.}{2019b}]{Jacobs_DES2}
Jacobs C.,  et~al., 2019b, \mn@doi [Monthly Notices of the Royal Astronomical
  Society] {10.1093/mnras/stz272}, 484, 5330

\bibitem[\protect\citeauthoryear{Jordan, Ghahramani, Jaakkola  \& Saul}{Jordan
  et~al.}{1999}]{Jordan_Variational}
Jordan M.~I.,  Ghahramani Z.,  Jaakkola T.~S.,   Saul L.~K.,  1999, An
  Introduction to Variational Methods for Graphical Models.
MIT Press, Cambridge, MA, USA, p. 105–161

\bibitem[\protect\citeauthoryear{{Jullo}, {Kneib}, {Limousin},
  {El{\'\i}asd{\'o}ttir}, {Marshall}  \& {Verdugo}}{{Jullo}
  et~al.}{2007}]{Jullo_2007}
{Jullo} E.,  {Kneib} J.~P.,  {Limousin} M.,  {El{\'\i}asd{\'o}ttir} {\'A}.,
  {Marshall} P.~J.,   {Verdugo} T.,  2007, \mn@doi [New Journal of Physics]
  {10.1088/1367-2630/9/12/447}, \href
  {https://ui.adsabs.harvard.edu/abs/2007NJPh....9..447J} {9, 447}

\bibitem[\protect\citeauthoryear{Keeton, Kochanek  \& Seljak}{Keeton
  et~al.}{1997}]{Keeton_Sie}
Keeton C.~R.,  Kochanek C.~S.,   Seljak U.,  1997, \mn@doi [The Astrophysical
  Journal] {10.1086/304172}, \href
  {https://ui.adsabs.harvard.edu/abs/1997ApJ...482..604K} {482, 604}

\bibitem[\protect\citeauthoryear{{Kendall} \& {Gal}}{{Kendall} \&
  {Gal}}{2017}]{Kendall_BNN}
{Kendall} A.,  {Gal} Y.,  2017, arXiv e-prints, \href
  {https://ui.adsabs.harvard.edu/abs/2017arXiv170304977K} {p. arXiv:1703.04977}

\bibitem[\protect\citeauthoryear{{Kingma} \& {Ba}}{{Kingma} \&
  {Ba}}{2014}]{Kingma_Adam}
{Kingma} D.~P.,  {Ba} J.,  2014, arXiv e-prints, \href
  {https://ui.adsabs.harvard.edu/abs/2014arXiv1412.6980K} {p. arXiv:1412.6980}

\bibitem[\protect\citeauthoryear{{Koekemoer} et~al.,}{{Koekemoer}
  et~al.}{2007}]{Koekemoer_2007}
{Koekemoer} A.~M.,  et~al., 2007, \mn@doi [\apjs] {10.1086/520086}, \href
  {https://ui.adsabs.harvard.edu/abs/2007ApJS..172..196K} {172, 196}

\bibitem[\protect\citeauthoryear{Koopmans}{Koopmans}{2005}]{Koopmans_DM}
Koopmans L. V.~E.,  2005, \mn@doi [Monthly Notices of the Royal Astronomical
  Society] {10.1111/j.1365-2966.2005.09523.x}, 363, 1136

\bibitem[\protect\citeauthoryear{{Koopmans}, {Treu}, {Bolton}, {Burles}  \&
  {Moustakas}}{{Koopmans} et~al.}{2006}]{Koopmans_06}
{Koopmans} L. V.~E.,  {Treu} T.,  {Bolton} A.~S.,  {Burles} S.,   {Moustakas}
  L.~A.,  2006, \mn@doi [\apj] {10.1086/505696}, \href
  {https://ui.adsabs.harvard.edu/abs/2006ApJ...649..599K} {649, 599}

\bibitem[\protect\citeauthoryear{{Kormann}, {Schneider}  \&
  {Bartelmann}}{{Kormann} et~al.}{1994}]{Kormann_SIE}
{Kormann} R.,  {Schneider} P.,   {Bartelmann} M.,  1994, \aap, \href
  {https://ui.adsabs.harvard.edu/abs/1994A&A...284..285K} {284, 285}

\bibitem[\protect\citeauthoryear{Kull, Filho  \& Flach}{Kull
  et~al.}{2017}]{Kull_BetaCalibration}
Kull M.,  Filho T.~S.,   Flach P.,  2017, in Singh A.,  Zhu J.,  eds,
  Proceedings of Machine Learning Research Vol. 54, Proceedings of the 20th
  International Conference on Artificial Intelligence and Statistics. PMLR, pp
  623--631, \url {https://proceedings.mlr.press/v54/kull17a.html}

\bibitem[\protect\citeauthoryear{Kullback \& Leibler}{Kullback \&
  Leibler}{1951}]{Kullback_Divergence}
Kullback S.,  Leibler R.~A.,  1951, The annals of mathematical statistics, 22,
  79

\bibitem[\protect\citeauthoryear{{LSST Science Collaboration} et~al.,}{{LSST
  Science Collaboration} et~al.}{2009}]{LSST}
{LSST Science Collaboration} et~al., 2009, arXiv e-prints, \href
  {https://ui.adsabs.harvard.edu/abs/2009arXiv0912.0201L} {p. arXiv:0912.0201}

\bibitem[\protect\citeauthoryear{Laureijs et~al.,}{Laureijs
  et~al.}{2011}]{Laureijs_Euclid}
Laureijs R.,  et~al., 2011, arXiv e-prints, \href
  {https://ui.adsabs.harvard.edu/abs/2011arXiv1110.3193L} {p. arXiv:1110.3193}

\bibitem[\protect\citeauthoryear{LeCun, Bengio  \& Hinton}{LeCun
  et~al.}{2015}]{LeCun_CNNReview}
LeCun Y.,  Bengio Y.,   Hinton G.,  2015, \mn@doi [Nature]
  {10.1038/nature14539}, 521, 436

\bibitem[\protect\citeauthoryear{{Lefor}, {Futamase}  \& {Akhlaghi}}{{Lefor}
  et~al.}{2013}]{Lefor_2013}
{Lefor} A.~T.,  {Futamase} T.,   {Akhlaghi} M.,  2013, \mn@doi [\nar]
  {10.1016/j.newar.2013.05.001}, \href
  {https://ui.adsabs.harvard.edu/abs/2013NewAR..57....1L} {57, 1}

\bibitem[\protect\citeauthoryear{Li et~al.,}{Li et~al.}{2020}]{Li_KiDS}
Li R.,  et~al., 2020, \mn@doi [The Astrophysical Journal]
  {10.3847/1538-4357/ab9dfa}, \href
  {https://ui.adsabs.harvard.edu/abs/2020ApJ...899...30L} {899, 30}

\bibitem[\protect\citeauthoryear{Mao \& Schneider}{Mao \&
  Schneider}{1998}]{mao_substructure}
Mao S.,  Schneider P.,  1998, \mn@doi [Monthly Notices of the Royal
  Astronomical Society] {10.1046/j.1365-8711.1998.01319.x}, 295, 587

\bibitem[\protect\citeauthoryear{{Massey}, {Stoughton}, {Leauthaud}, {Rhodes},
  {Koekemoer}, {Ellis}  \& {Shaghoulian}}{{Massey} et~al.}{2010}]{Massey_2010}
{Massey} R.,  {Stoughton} C.,  {Leauthaud} A.,  {Rhodes} J.,  {Koekemoer} A.,
  {Ellis} R.,   {Shaghoulian} E.,  2010, \mn@doi [\mnras]
  {10.1111/j.1365-2966.2009.15638.x}, \href
  {https://ui.adsabs.harvard.edu/abs/2010MNRAS.401..371M} {401, 371}

\bibitem[\protect\citeauthoryear{{Metcalf} et~al.,}{{Metcalf}
  et~al.}{2019}]{Metcalf_Challenge}
{Metcalf} R.~B.,  et~al., 2019, \mn@doi [\aap] {10.1051/0004-6361/201832797},
  \href {https://ui.adsabs.harvard.edu/abs/2019A&A...625A.119M} {625, A119}

\bibitem[\protect\citeauthoryear{{Napolitano} et~al.,}{{Napolitano}
  et~al.}{2020}]{Napolitano_2020}
{Napolitano} N.~R.,  et~al., 2020, \mn@doi [\apjl] {10.3847/2041-8213/abc95b},
  \href {https://ui.adsabs.harvard.edu/abs/2020ApJ...904L..31N} {904, L31}

\bibitem[\protect\citeauthoryear{Nightingale, Dye  \& Massey}{Nightingale
  et~al.}{2018a}]{nightingale_Autolens}
Nightingale J.~W.,  Dye S.,   Massey R.~J.,  2018a, \mn@doi [Monthly Notices of
  the Royal Astronomical Society] {10.1093/mnras/sty1264}, 478, 4738

\bibitem[\protect\citeauthoryear{{Nightingale}, {Dye}  \&
  {Massey}}{{Nightingale} et~al.}{2018b}]{Nightingale_2018}
{Nightingale} J.~W.,  {Dye} S.,   {Massey} R.~J.,  2018b, \mn@doi [\mnras]
  {10.1093/mnras/sty1264}, \href
  {https://ui.adsabs.harvard.edu/abs/2018MNRAS.478.4738N} {478, 4738}

\bibitem[\protect\citeauthoryear{{Oguri} et~al.,}{{Oguri}
  et~al.}{2006}]{Oguri_LRG}
{Oguri} M.,  et~al., 2006, \mn@doi [\aj] {10.1086/506019}, \href
  {https://ui.adsabs.harvard.edu/abs/2006AJ....132..999O} {132, 999}

\bibitem[\protect\citeauthoryear{Pearson, Li  \& Dye}{Pearson
  et~al.}{2019}]{pearson_ModellingCNN}
Pearson J.,  Li N.,   Dye S.,  2019, \mn@doi [Monthly Notices of the Royal
  Astronomical Society] {10.1093/mnras/stz1750}, 488, 991

\bibitem[\protect\citeauthoryear{Pearson, Maresca, Li  \& Dye}{Pearson
  et~al.}{2021}]{pearson_CombinedCBNN}
Pearson J.,  Maresca J.,  Li N.,   Dye S.,  2021, arXiv e-prints, 2103,
  arXiv:2103.03257

\bibitem[\protect\citeauthoryear{Perreault~Levasseur, Hezaveh  \&
  Wechsler}{Perreault~Levasseur et~al.}{2017}]{Levasseur_modellingBNN}
Perreault~Levasseur L.,  Hezaveh Y.~D.,   Wechsler R.~H.,  2017, \mn@doi [The
  Astrophysical Journal Letters] {10.3847/2041-8213/aa9704}, 850, L7

\bibitem[\protect\citeauthoryear{Petrillo et~al.,}{Petrillo
  et~al.}{2017}]{Petrillo_1}
Petrillo C.~E.,  et~al., 2017, \mn@doi [Monthly Notices of the Royal
  Astronomical Society] {10.1093/mnras/stx2052}, \href
  {https://ui.adsabs.harvard.edu/abs/2017MNRAS.472.1129P} {472, 1129}

\bibitem[\protect\citeauthoryear{Petrillo et~al.,}{Petrillo
  et~al.}{2019a}]{Petrillo_2}
Petrillo C.~E.,  et~al., 2019a, \mn@doi [Monthly Notices of the Royal
  Astronomical Society] {10.1093/mnras/sty2683}, \href
  {https://ui.adsabs.harvard.edu/abs/2019MNRAS.482..807P} {482, 807}

\bibitem[\protect\citeauthoryear{Petrillo et~al.,}{Petrillo
  et~al.}{2019b}]{Petrillo_3}
Petrillo C.~E.,  et~al., 2019b, \mn@doi [Monthly Notices of the Royal
  Astronomical Society] {10.1093/mnras/stz189}, \href
  {https://ui.adsabs.harvard.edu/abs/2019MNRAS.484.3879P} {484, 3879}

\bibitem[\protect\citeauthoryear{{Pourrahmani}, {Nayyeri}  \&
  {Cooray}}{{Pourrahmani} et~al.}{2018}]{Pourrahmani_LensFlow}
{Pourrahmani} M.,  {Nayyeri} H.,   {Cooray} A.,  2018, \mn@doi [\apj]
  {10.3847/1538-4357/aaae6a}, \href
  {https://ui.adsabs.harvard.edu/abs/2018ApJ...856...68P} {856, 68}

\bibitem[\protect\citeauthoryear{Refsdal}{Refsdal}{1964}]{refsdal_hubble}
Refsdal S.,  1964, \mn@doi [Monthly Notices of the Royal Astronomical Society]
  {10.1093/mnras/128.4.307}, 128, 307

\bibitem[\protect\citeauthoryear{{Ryon}}{{Ryon}}{2022}]{Ryon_ACS}
{Ryon} J.~E.,  2022, in , Vol.~21, ACS Instrument Handbook for Cycle 30 v.
  21.0.
p.~21

\bibitem[\protect\citeauthoryear{{Scaramella} et~al.,}{{Scaramella}
  et~al.}{2021}]{Scaramella_Euclid_WS}
{Scaramella} R.,  et~al., 2021, arXiv e-prints, \href
  {https://ui.adsabs.harvard.edu/abs/2021arXiv210801201S} {p. arXiv:2108.01201}

\bibitem[\protect\citeauthoryear{Schneider, Ehlers  \& Falco}{Schneider
  et~al.}{1992}]{schneider_1992}
Schneider P.,  Ehlers J.,   Falco E.~E.,  1992, Gravitational Lenses.
{Springer-Verlag New York}, \mn@doi{10.1007/978-3-662-03758-4}

\bibitem[\protect\citeauthoryear{{Schuldt}, {Suyu}, {Meinhardt},
  {Leal-Taix{\'e}}, {Ca{\~n}ameras}, {Taubenberger}  \& {Halkola}}{{Schuldt}
  et~al.}{2021}]{Schuldt_ModellingCNN}
{Schuldt} S.,  {Suyu} S.~H.,  {Meinhardt} T.,  {Leal-Taix{\'e}} L.,
  {Ca{\~n}ameras} R.,  {Taubenberger} S.,   {Halkola} A.,  2021, \mn@doi [\aap]
  {10.1051/0004-6361/202039574}, \href
  {https://ui.adsabs.harvard.edu/abs/2021A&A...646A.126S} {646, A126}

\bibitem[\protect\citeauthoryear{{Schuldt}, {Ca{\~n}ameras}, {Shu}, {Suyu},
  {Taubenberger}, {Meinhardt}  \& {Leal-Taix{\'e}}}{{Schuldt}
  et~al.}{2022}]{Schuldt_22}
{Schuldt} S.,  {Ca{\~n}ameras} R.,  {Shu} Y.,  {Suyu} S.~H.,  {Taubenberger}
  S.,  {Meinhardt} T.,   {Leal-Taix{\'e}} L.,  2022, arXiv e-prints, \href
  {https://ui.adsabs.harvard.edu/abs/2022arXiv220611279S} {p. arXiv:2206.11279}

\bibitem[\protect\citeauthoryear{Serjeant}{Serjeant}{2014}]{Serjeant_Lenses}
Serjeant S.,  2014, \mn@doi [The Astrophysical Journal Letters]
  {10.1088/2041-8205/793/1/L10}, 793, L10

\bibitem[\protect\citeauthoryear{S{\'e}rsic}{S{\'e}rsic}{1963}]{Sersic_Profile}
S{\'e}rsic J.~L.,  1963, Boletin de la Asociacion Argentina de Astronomia La
  Plata Argentina, \href
  {https://ui.adsabs.harvard.edu/abs/1963BAAA....6...41S} {6, 41}

\bibitem[\protect\citeauthoryear{{Shajib}, {Treu}, {Birrer}  \&
  {Sonnenfeld}}{{Shajib} et~al.}{2021}]{Shajib_21}
{Shajib} A.~J.,  {Treu} T.,  {Birrer} S.,   {Sonnenfeld} A.,  2021, \mn@doi
  [\mnras] {10.1093/mnras/stab536}, \href
  {https://ui.adsabs.harvard.edu/abs/2021MNRAS.503.2380S} {503, 2380}

\bibitem[\protect\citeauthoryear{{Shu} et~al.,}{{Shu}
  et~al.}{2017a}]{Slacs_XIII}
{Shu} Y.,  et~al., 2017a, \mn@doi [\apj] {10.3847/1538-4357/aa9794}, \href
  {https://ui.adsabs.harvard.edu/abs/2017ApJ...851...48S} {851, 48}

\bibitem[\protect\citeauthoryear{{Shu} et~al.,}{{Shu} et~al.}{2017b}]{SLACS_13}
{Shu} Y.,  et~al., 2017b, \mn@doi [\apj] {10.3847/1538-4357/aa9794}, \href
  {https://ui.adsabs.harvard.edu/abs/2017ApJ...851...48S} {851, 48}

\bibitem[\protect\citeauthoryear{{Sonnenfeld}}{{Sonnenfeld}}{2021a}]{sonnenfeld_2021c}
{Sonnenfeld} A.,  2021a, arXiv e-prints, \href
  {https://ui.adsabs.harvard.edu/abs/2021arXiv210913246S} {p. arXiv:2109.13246}

\bibitem[\protect\citeauthoryear{{Sonnenfeld}}{{Sonnenfeld}}{2021b}]{sonnenfeld_2021d}
{Sonnenfeld} A.,  2021b, arXiv e-prints, \href
  {https://ui.adsabs.harvard.edu/abs/2021arXiv211009537S} {p. arXiv:2110.09537}

\bibitem[\protect\citeauthoryear{{Sonnenfeld}}{{Sonnenfeld}}{2021c}]{sonnenfeld_2021b}
{Sonnenfeld} A.,  2021c, \mn@doi [\aap] {10.1051/0004-6361/202142062}, \href
  {https://ui.adsabs.harvard.edu/abs/2021A&A...656A.153S} {656, A153}

\bibitem[\protect\citeauthoryear{{Sonnenfeld} \& {Cautun}}{{Sonnenfeld} \&
  {Cautun}}{2021}]{sonnenfeld_2021a}
{Sonnenfeld} A.,  {Cautun} M.,  2021, \mn@doi [\aap]
  {10.1051/0004-6361/202140549}, \href
  {https://ui.adsabs.harvard.edu/abs/2021A&A...651A..18S} {651, A18}

\bibitem[\protect\citeauthoryear{{Sonnenfeld}, {Gavazzi}, {Suyu}, {Treu}  \&
  {Marshall}}{{Sonnenfeld} et~al.}{2013}]{Sonnenfeld_SL2S}
{Sonnenfeld} A.,  {Gavazzi} R.,  {Suyu} S.~H.,  {Treu} T.,   {Marshall} P.~J.,
  2013, \mn@doi [\apj] {10.1088/0004-637X/777/2/97}, \href
  {https://ui.adsabs.harvard.edu/abs/2013ApJ...777...97S} {777, 97}

\bibitem[\protect\citeauthoryear{{Sonnenfeld}, {Treu}, {Marshall}, {Suyu},
  {Gavazzi}, {Auger}  \& {Nipoti}}{{Sonnenfeld} et~al.}{2015a}]{Sonnenfeld_15}
{Sonnenfeld} A.,  {Treu} T.,  {Marshall} P.~J.,  {Suyu} S.~H.,  {Gavazzi} R.,
  {Auger} M.~W.,   {Nipoti} C.,  2015a, \mn@doi [\apj]
  {10.1088/0004-637X/800/2/94}, \href
  {https://ui.adsabs.harvard.edu/abs/2015ApJ...800...94S} {800, 94}

\bibitem[\protect\citeauthoryear{{Sonnenfeld}, {Treu}, {Marshall}, {Suyu},
  {Gavazzi}, {Auger}  \& {Nipoti}}{{Sonnenfeld} et~al.}{2015b}]{SL2S_last}
{Sonnenfeld} A.,  {Treu} T.,  {Marshall} P.~J.,  {Suyu} S.~H.,  {Gavazzi} R.,
  {Auger} M.~W.,   {Nipoti} C.,  2015b, \mn@doi [\apj]
  {10.1088/0004-637X/800/2/94}, \href
  {https://ui.adsabs.harvard.edu/abs/2015ApJ...800...94S} {800, 94}

\bibitem[\protect\citeauthoryear{Sonnenfeld, Jaelani, Chan, More, Suyu, Wong,
  Oguri  \& Lee}{Sonnenfeld et~al.}{2019}]{Sonnenfeld_IMF}
Sonnenfeld A.,  Jaelani A.~T.,  Chan J.,  More A.,  Suyu S.~H.,  Wong K.~C.,
  Oguri M.,   Lee C.-H.,  2019, \mn@doi [Astronomy \& Astrophysics]
  {10.1051/0004-6361/201935743}, 630, A71

\bibitem[\protect\citeauthoryear{Spiniello, Koopmans, Trager, Czoske  \&
  Treu}{Spiniello et~al.}{2011}]{spiniello_DM}
Spiniello C.,  Koopmans L. V.~E.,  Trager S.~C.,  Czoske O.,   Treu T.,  2011,
  \mn@doi [Monthly Notices of the Royal Astronomical Society]
  {10.1111/j.1365-2966.2011.19458.x}, 417, 3000

\bibitem[\protect\citeauthoryear{Srivastava, Hinton, Krizhevsky, Sutskever  \&
  Salakhutdinov}{Srivastava et~al.}{2014}]{Srivastava_Dropout}
Srivastava N.,  Hinton G.,  Krizhevsky A.,  Sutskever I.,   Salakhutdinov R.,
  2014, Journal of Machine Learning Research, 15, 1929

\bibitem[\protect\citeauthoryear{Tortora, Napolitano, Romanowsky  \&
  Jetzer}{Tortora et~al.}{2010}]{Tortora_DM}
Tortora C.,  Napolitano N.~R.,  Romanowsky A.~J.,   Jetzer P.,  2010, \mn@doi
  [The Astrophysical Journal Letters] {10.1088/2041-8205/721/1/L1}, 721, L1

\bibitem[\protect\citeauthoryear{Treu}{Treu}{2010}]{Treu_Lensing}
Treu T.,  2010, \mn@doi [Annual Review of Astronomy and Astrophysics]
  {10.1146/annurev-astro-081309-130924}, \href
  {https://ui.adsabs.harvard.edu/abs/2010ARA\&A..48...87T} {48, 87}

\bibitem[\protect\citeauthoryear{Treu \& Koopmans}{Treu \&
  Koopmans}{2004}]{treu_DM}
Treu T.,  Koopmans L. V.~E.,  2004, \mn@doi [The Astrophysical Journal]
  {10.1086/422245}, 611, 739

\bibitem[\protect\citeauthoryear{Treu, Auger, Koopmans, Gavazzi, Marshall  \&
  Bolton}{Treu et~al.}{2010}]{Treu_IMF}
Treu T.,  Auger M.~W.,  Koopmans L. V.~E.,  Gavazzi R.,  Marshall P.~J.,
  Bolton A.~S.,  2010, \mn@doi [ApJ] {10.1088/0004-637X/709/2/1195}, 709, 1195

\bibitem[\protect\citeauthoryear{{Treu}, {Dutton}, {Auger}, {Marshall},
  {Bolton}, {Brewer}, {Koo}  \& {Koopmans}}{{Treu} et~al.}{2011}]{Treu_SWELLS}
{Treu} T.,  {Dutton} A.~A.,  {Auger} M.~W.,  {Marshall} P.~J.,  {Bolton} A.~S.,
   {Brewer} B.~J.,  {Koo} D.~C.,   {Koopmans} L. V.~E.,  2011, \mn@doi [\mnras]
  {10.1111/j.1365-2966.2011.19378.x}, \href
  {https://ui.adsabs.harvard.edu/abs/2011MNRAS.417.1601T} {417, 1601}

\bibitem[\protect\citeauthoryear{{Vegetti} \& {Koopmans}}{{Vegetti} \&
  {Koopmans}}{2009}]{Vegetti_2008}
{Vegetti} S.,  {Koopmans} L.~V.~E.,  2009, \mn@doi [\mnras]
  {10.1111/j.1365-2966.2008.14005.x}, \href
  {https://ui.adsabs.harvard.edu/abs/2009MNRAS.392..945V} {392, 945}

\bibitem[\protect\citeauthoryear{Vegetti, Koopmans, Auger, Treu  \&
  Bolton}{Vegetti et~al.}{2014}]{Vegetti_DM}
Vegetti S.,  Koopmans L. V.~E.,  Auger M.~W.,  Treu T.,   Bolton A.~S.,  2014,
  \mn@doi [Monthly Notices of the Royal Astronomical Society]
  {10.1093/mnras/stu943}, 442, 2017

\bibitem[\protect\citeauthoryear{{Vernardos} \& {Koopmans}}{{Vernardos} \&
  {Koopmans}}{2022}]{Vernardos_22}
{Vernardos} G.,  {Koopmans} L.~V.~E.,  2022, \mn@doi [\mnras]
  {10.1093/mnras/stac1924}, \href
  {https://ui.adsabs.harvard.edu/abs/2022MNRAS.516.1347V} {516, 1347}

\bibitem[\protect\citeauthoryear{Virtanen et~al.,}{Virtanen
  et~al.}{2020}]{Virtanen_SciPy}
Virtanen P.,  et~al., 2020, \mn@doi [Nature Methods]
  {10.1038/s41592-019-0686-2}, \href {https://rdcu.be/b08Wh} {17, 261}

\bibitem[\protect\citeauthoryear{{Weaver} et~al.,}{{Weaver}
  et~al.}{2022}]{Weaver_2021}
{Weaver} J.~R.,  et~al., 2022, \mn@doi [\apjs] {10.3847/1538-4365/ac3078},
  \href {https://ui.adsabs.harvard.edu/abs/2022ApJS..258...11W} {258, 11}

\bibitem[\protect\citeauthoryear{Wong et~al.,}{Wong
  et~al.}{2020}]{Wong_Holicow2}
Wong K.~C.,  et~al., 2020, \mn@doi [Monthly Notices of the Royal Astronomical
  Society] {10.1093/mnras/stz3094}, \href
  {https://ui.adsabs.harvard.edu/abs/2020MNRAS.498.1420W} {498, 1420}

\bibitem[\protect\citeauthoryear{Zadrozny \& Elkan}{Zadrozny \&
  Elkan}{2001}]{Zadrozny_Calibration}
Zadrozny B.,  Elkan C.,  2001, in Proceedings of the Eighteenth International
  Conference on Machine Learning. ICML '01.
Morgan Kaufmann Publishers Inc., San Francisco, CA, USA, p. 609–616

\bibitem[\protect\citeauthoryear{Zadrozny \& Elkan}{Zadrozny \&
  Elkan}{2002}]{Zadrozny_Calibration2}
Zadrozny B.,  Elkan C.,  2002, in Proceedings of the Eighth ACM SIGKDD
  International Conference on Knowledge Discovery and Data Mining. KDD '02.
Association for Computing Machinery, New York, NY, USA, p. 694–699,
  \mn@doi{10.1145/775047.775151}, \url {https://doi.org/10.1145/775047.775151}

\bibitem[\protect\citeauthoryear{Zwicky}{Zwicky}{1937}]{Zwicky_Lensing}
Zwicky F.,  1937, \mn@doi [Physical Review] {10.1103/PhysRev.51.290}, \href
  {https://ui.adsabs.harvard.edu/abs/1937PhRv...51..290Z} {51, 290}

\bibitem[\protect\citeauthoryear{{de Vaucouleurs}}{{de
  Vaucouleurs}}{1948}]{deVaucouleurs_Profile}
{de Vaucouleurs} G.,  1948, Annales d'Astrophysique, \href
  {https://ui.adsabs.harvard.edu/abs/1948AnAp...11..247D} {11, 247}

\makeatother
\end{thebibliography}

\bsp	
\label{lastpage}
\end{document}